\documentclass[epsfig,12pt]{article}
\usepackage{epsfig}
\usepackage{graphicx}
\usepackage{float,subcaption}
\usepackage{array}
\usepackage{color}
\usepackage{bm}
\usepackage{bbm}
\usepackage{amssymb,amsmath,mathtools,physics}
\numberwithin{equation}{section}
\usepackage{cancel}
\usepackage{threeparttable}
\usepackage{adjustbox}
\usepackage{hyperref}
\hypersetup{
    colorlinks=true,
    allcolors = black
}
\usepackage[numbers,sort&compress]{natbib}
\usepackage[nottoc,notlof,notlot]{tocbibind} 
\usepackage[titles]{tocloft} 
\usepackage{booktabs}
\usepackage[T1]{fontenc}

\usepackage[vmargin={1.2in,1in}]{geometry}

\newcommand{\beq}{\begin{equation}}   
\newcommand{\eeq}{\end{equation}}
\newcommand{\beqn}{\begin{eqnarray}}   
\newcommand{\eeqn}{\end{eqnarray}}

\newcommand{\pt}{\partial}

\def\ntwot{${\mathcal N}=(2,2)$}
\def\ntwoo{${\mathcal N}=(0,2)$}

\newcommand{\gsim}{\lower.7ex\hbox{$
\;\stackrel{\textstyle>}{\sim}\;$}}
\newcommand{\lsim}{\lower.7ex\hbox{$
\;\stackrel{\textstyle<}{\sim}\;$}}
\def\bar{\overline}
\def\tilde{\widetilde}
\def\d{\partial}
\def\cp{\mathbb{CP}}
\def\bphi{\overline{\varphi}}
\def\bpsi{\overline{\psi}}
\def\bq{\overline{q}}
\def\rie{R_{1\bar{1}1\bar{1}}}

\def\NN{\mathbb{N}}
\def\RR{\mathbb{R}}
\def\ZZ{\mathbb{Z}}

\def\cL{\mathcal{L}}

\def\cR{\mathcal{R}}
\def\cT{\mathcal{T}}

\def\cJ{\mathcal{J}}

\DeclareMathOperator{\arctanh}{arctanh}
\DeclareMathOperator{\arccosh}{arccosh}
\DeclareMathOperator{\sn}{sn}
\DeclareMathOperator{\cn}{cn}
\DeclareMathOperator{\sd}{sd}
\DeclareMathOperator{\cd}{cd}
\DeclareMathOperator{\dn}{dn}

\newcommand\Tstrut{\rule{0pt}{2.6ex}}         % = `top' strut
\newcommand\Bstrut{\rule[-0.9ex]{0pt}{0pt}}   % = `bottom' strut

\usepackage[textsize=scriptsize]{todonotes}

\begin{document}
\unitlength = 1mm

\begin{titlepage}

   \begin{flushright}
    FTPI-MINN-24-05, UMN-TH-4314/24\\
    \end{flushright}

\begin{center}
{\Large \bf Lie-algebraic K\"ahler sigma models\\[2mm]
with the U(1) isotropy }

\end{center}

 \vspace{5mm}
    
    \begin{center}
    { \bf   Chao-Hsiang Sheu$^{a}$ and Mikhail Shifman$^{a, b}$}
    \end {center}
    
    \begin{center}
    
        {\it  $^{a}$Department of Physics,
    University of Minnesota,
    Minneapolis, MN 55455}\\{\small and}\\
    {\it  $^{b}$William I. Fine Theoretical Physics Institute,
    University of Minnesota,
    Minneapolis, MN 55455}\\
    
    \end{center}

\vspace{10mm}

\begin{center}
{\bf Abstract}
\end{center} 
We discuss various questions which emerge in connection with the Lie-algebraic deformation of $\cp^1$ sigma model in two dimensions. First we supersymmetrize the original model endowing it with the minimal ${\cal N}=(0,2)$ and extended ${\cal N}=(2,2)$ supersymmetries. Then we derive the general hypercurrent anomaly 
in the both cases. In the latter case this anomaly is one-loop but is somewhat different from the standard expressions one can find in the literature because the target manifold is non-symmetric. We also show how to introduce the twisted masses and the $\theta$ term, and study the BPS equation for instantons,
in particular the value of the topological charge. 
Then we demonstrate that the second  loop in the $\beta$ function of the {\em non}-supersymmetric Lie-algebraic sigma model is due to an infrared effect. To this end we use a {\em supersymmetric} regularization. We also conjecture that the above statement is valid for higher loops too, similar to the parallel phenomenon in four-dimensional ${\cal N}=1$ super-Yang-Mills. In the second part of the paper we develop a special dimensional reduction -- namely, starting from the two-dimensional Lie-algebraic model we arrive at a quasi-exactly solvable quantum-mechanical problem of the Lam\'e type.

\end{titlepage}

\tableofcontents
\newpage

\section{Introduction}

In this paper  we continue the studies of one(complex)-dimensional sigma models on K\"ahlerian target spaces which generalize the $\cp^1$ model in a Lie-algebraic way \cite{g1,g2,g3}.
For practical applications in baby Skyrmions this model is usually formulated in the from
\beq
{\mathcal L} =\frac{1}{2g^2(S_3)} (\pt S_i) (\pt S_i)\,,\qquad \vec{S} \vec{S}=1\,.
\label{1one}
\eeq
where the coupling $g^2$ becomes a function of $S_3$, the third component of the isovector $\vec S$,
\beq
g^2(S_3) = g^2\cdot  \left(
\textstyle{\frac{1+k}{2}}+\textstyle{\frac{1-k}{2}}\,S_3^2\right) .
\label{2two}
\eeq
Moreover,  $k$ is a numerical parameter  defined below in Eq. (\ref{5five}). At $k=1$ we return to the Heisenberg O(3) model. With $k\neq 1$
the round metric is deformed. 

For theoretical applications in two dimensions (2D)
it is more convenient to use the geometric representation,
\beq
{\mathcal L} =G_{1\bar 1}\left(\pt_\mu\bar\varphi \pt^\mu\varphi\right),
\label{3three}
\eeq
where $G_{1\bar 1}$ is a generalization of the Fubini-Study metric,\footnote{The equality $n_1= n_3$ can always be achieved by rescaling 
the fields $\varphi, \bar\varphi$.}
\beqn
G_{1\bar 1} &=& \frac{1}{n_1 +n_2\bar\varphi \varphi + n_3 (\bar\varphi\varphi)^2}\,,
\label{4four} \\[2mm]
n_1&=& n_3= \frac{g^2}{2}, \quad  n_2 = g^2k\,.
\label{5five}
\eeqn
If $k=1$ the metric (\ref{4four}) is the standard Fubini-Study metric. In what follows we will use a simplified notation,
\beq
G_{1\bar 1} \equiv G\,,\quad G^{1\bar 1} \equiv G^{-1}\,. \nonumber
\eeq
Other abbreviations are introduced in Eqs. (\ref{chris}) and (\ref{scal}).

\begin{figure}[t] 
  \centering 
  \includegraphics[width=.7\linewidth]{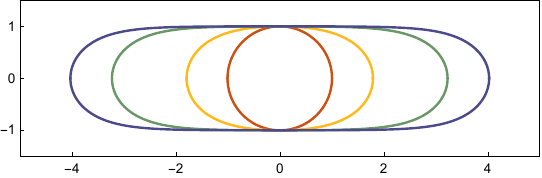}
  \caption{\small The orientation of $\vec{S}$-vector embedding in a three-Euclidean space. The contours from inside to outside (i.e. red to blue) correspond to $k$ equal to 1.0, 9.5, 200, and 1000, respectively.}
  \label{sausage}
\end{figure}

One can consider another deformation of this model, by the so-called twisted mass term \cite{Alvarez-Gaume:1983uye,Shifman:2006bs}.
Then, Eq. (\ref{3three}) takes the form
\beq
{\mathcal L}_m =G_{1\bar 1}\left(\pt_\mu\bar\varphi \pt^\mu\varphi -m^2\bar\varphi\varphi\right),
\label{6six}
\eeq
In temrs of representation (\ref{1one}) we then have
\beq
\cL = \frac{1}{2g^2(S_3)} \Bigl[(\pt S_i) (\pt S_i) - \abs{m}^2(1-S_3^2)\Bigr] \,.
\eeq
Both deformations (i.e. the metric deformation $k\neq 1$ and a ``potential'' deformation $m\neq 0$) destroy O(3) invariance of the target space and introduce a dependence on the orientation of $\vec{S}$ with respect to the third axis in the isospace, see Fig. \ref{sausage}.

Perturbation theory in the model at hand have been studied in \cite{g1,g2} in the framework of the so called first-order formalism related to an operator product expansion (OPE), see the list of references in \cite{g1,g2}. The following phenomenon have been observed there.

At one loop OPE for certain chiral currents have the form 
\beq
J_a(z) J_b (0) = \frac{1}{z} f^c_{ab} J_c(0)
\label{8eight}
\eeq
where $f^c_{ab}$ are the $sl(2)$ algebra structure constants.  The above expression  fully reveals the Lie-algebraic structure of (\ref{4four}).  However, a straightforward calculation of the second loop produces a term
\beq
\frac{1}{z^2}\pt_i v_a^j\pt_jv_b^i
\label{9nine}
\eeq
where $v_a^j$ are Killing vectors. The above structure is obviously non-geometric. However, one can show that  with a proper regularization within the first-order formalism the partial derivatives in (\ref{9nine}) are replaced by covariant derivatives, and 
the required ``geometricity'' is recovered. In Ref. \cite{g2} the regularization method was  based on supersymmetry despite the fact that
the calculated $\beta$ function was that of the non-supersymmetric model (\ref{3three}).
  In \cite{g1} the following $H$ hypothesis was formulated:  In (non-supersymmetric) K\"ahlerian sigma models, an anomaly is present in the calculation of the second and higher loops. In \cite{g2} the validity of the $H$ hypothesis  was verified in the second loop. 

In this paper we reveal an infrared anomaly in the second $\beta$-function coefficient. We use the standard perturbation theory and standard two-loop Feynman graphs. Our derivation is based on ${\cal N}=(2,2)$ supersymmetry; however, the strategy is different from that in \cite{g2}. Our analysis has close parallels with
the holonomy anomaly in ${\cal N}=1$ supersymmetric Yang-Mills (SYM) theory \cite{SV} and two-dimemsional
sigma models \cite{Cui:2011uw,Chen:2014efa}. Just like in the latter case it is likely that the anomalous effect detected in non-supersymmetric K\"ahlerian sigma models can be reformulated as a subtlety in the measure in the corresponding path integral \cite{AHM}.

Our work consist of several parts. First, we supersymmetrize the model (\ref{3three}), (\ref{4four}), presenting both ${\cal N}=(2,2)$ and ${\cal N}=(0,2)$
versions. 
Section \ref {sec:hypercurrent} is devoted to the study of the hypercurrents in ${\cal N}=(2,2)$ and ${\cal N}=(0,2)$ versions of the model. The standard expressions known in the literature have to be modified to take into account the non-symmetric nature of the target space.
Then, in Secs. \ref{sec:twistedmass} and \ref{sec:thetaterm} we introduce the twisted mass and the $\theta$ angle.

In Sec. \ref{sec:betafunc} we calculate the two-loop $\beta$ function coefficient in the non-supersymmet\-ric sigma model by virtue of a supersymmet\-ric regularization. Our calculation is transparent and demonstrates the role of the infrared contribution.  Section \ref{sec:lame} is devoted to an interesting aspect of reducing the two-dimensional model under consideration to a Lie-algebraic quantum-mechanical model \cite{turb,shifm} presenting the so-called Lam\'e problem \cite{dunne}. Under certain values of quantized free parameters it becomes quasi-exactly solvable, and, moreover, exhibits duality in the non-supersymmetric case.

\section{Extending the deformed \boldmath{$\cp^1$} model}
\label{sec:extension}

In this section, we elaborate on supersymmetric extensions of the deformed $\cp^1$ model, incorporating twisted masses and a topological term in a consistent manner. Applications of the present construction to the analysis of $\beta$ functions and the reduced quantum mechanical model can be found in Sec. \ref{sec:betafunc} and \ref{sec:lame}, respectively. 

\subsection{ \boldmath{\ntwot} and \boldmath{\ntwoo} supersymmetrization}
\label{sec:susy}

\subsubsection{ \boldmath{\ntwot}}
\label{211}

We start with a brief review of the general construction of two-dimensional {\ntwot} and {\ntwoo} sigma models. The target space of the model under consideration is a one (complex) dimensional manifold; it is K\"ahlerian and  admits {\ntwot} structure \cite{Alvarez-Gaume:1981exv}. Since the basics of the {\ntwot} model can be found in standard textbooks, we just quote the results and remind the relevant geometric data. Suppose the target space is parametrized by the complex coordinates $\varphi$ and $\bphi$. By promoting the scalar field to the corresponding superfields, $\Phi,\Phi^{\dagger}$, and integrating out the Grassmann coordinates, one finds the component formulation (see for example \cite{hkp,Perelomov:1989im}), namely,
\begin{align}\label{eq:L22}
	\cL_{(2,2)} = G\left[ 
        \d_{\mu}\varphi\d^{\mu}\bphi{} 
        + i \bpsi\cancel{\d}\psi 
        + i\bpsi\gamma^{\mu}(\Gamma\d_{\mu}\varphi)\psi
    \right]
    - \frac{1}{2}\rie\left( \bpsi{}\psi \right)^2
\end{align} 
where  $\psi$ is a Dirac fermion, 
\begin{align}
	\psi = \mqty(\psi_R \\ \psi_L) 
	\,,\quad
	\bpsi = \psi^{\dagger}\gamma^0
	\,,
\end{align}
and
\beq
\Gamma\equiv \Gamma^1_{11}\label{chris}
\eeq
is the Christoffel symbol.
The essential geometric data (in addition to Eq. (\ref{4four})) are
\begin{subequations}\label{eq:23}
\begin{align}\label{eq:metric}
	&\Gamma 
	= -\frac{\bphi{}(n_2+2n_3\abs{\varphi}^2)}{n_1+n_2\abs{\varphi}^2+n_3\abs{\varphi}^4}
;	\\[3mm]
	\label{eq:rie}
	&\rie  = -\frac{1}{2}G^2\cR
	= -\frac{n_1n_2+4n_1n_3\abs{\varphi}^2+n_2n_3\abs{\varphi}^4}{(n_1 + n_2 \abs{\varphi}^2 + n_3\abs{\varphi}^4)^3}
\end{align}
\end{subequations}
in which $\cR$ is the scalar curvature,
\beq
{\cal R} = 2 G^{\bar n m} R_{m\bar n}\,, \quad R_{m\bar n} = -G^{\bar j i} R_{i\bar j m\bar n}\,.
\label{scal}
\eeq

\subsubsection{ \boldmath{\ntwoo}}
\label{212}

\noindent
As for the {\ntwoo} formulation, we limit our discussion to the so-called minimal model \cite{Cui:2011uw}. Following  the same lines as in \cite{Witten:2005px,Chen:2014efa,Shifman:2008wv}, we introduce an {\ntwoo} chiral superfield $A$ which, in terms of component fields, 
takes the form
\begin{align}
	A(x,\theta,\theta^{\dagger}) = \varphi(x) + \sqrt{2}\theta\psi_L(x) + i\theta^{\dagger}\theta \d_L\varphi(x)
\end{align}
where $\d_L = \d_{t} + \d_{z}$ and $\theta,\theta^{\dagger}$ are the  Grassmann coordinates. The $(0,2)$ supersymmetric transformation of $A$ is
\begin{align}
	\delta_{\epsilon,\epsilon^{\dagger}} A = \d_L\varphi \cdot 2i\epsilon^{\dagger}\theta + \sqrt{2}\epsilon\psi_L \,.
\end{align}
The Lagrangian can be written as 
\begin{align}\label{eq:L02}
	\cL_{(0,2)} &= \frac{1}{4} \int\dd^{2}{\theta} \left[ K_{1}(A,A^{\dagger})i\d_R A + \mbox{h.c.} \right]
	\nonumber\\[2mm]
	&= G\left[ 
        \d_{\mu}\varphi\d^{\mu}\bphi
        + i \bpsi_L\d_R\psi_L
        + i\bpsi_L(\Gamma \d_{R}\varphi)\psi_L
    \right].
\end{align}
Note that $K_1$ is the first derivative of the K\"ahler potential,\footnote{The explicit form of the K\"ahler potential $K$ of the deformed $\cp^1$ model is given in \cite{g3}.}
\begin{align}
	K_1 \equiv \pdv{K}{A} = \frac{2}{A\sqrt{n_2^2-4n_1n_3}}
	\arctanh\left(  
		\frac{AA^{\dagger}\sqrt{n_2^2-4n_1n_3}}{2n_1+n_2 AA^{\dagger}}
	\right).
\end{align}
By construction, Eq. \eqref{eq:L02} is {\ntwoo} invariant.
In contrast to the undeformed model, there is no nonlinear transformation of $A$ corresponding to the global rotations other than the $U(1)$ which can be straightforwardly seen in the above formulation.

As a side remark, we note that the {\ntwot} supersymmetry can be recovered from Eq. \eqref{eq:L02} by introduced another {\ntwoo} superfield $B$,
\begin{align}
	B(x,\theta,\theta^{\dagger}) &= \psi_R(x) + \sqrt{2}\theta F(x) + i \theta^{\dagger}\theta\d_L\psi_R(x)
\end{align}
obeying the transformation 
\begin{align}
	\delta_{\epsilon,\epsilon^{\dagger}} B = \d_L\psi_R \cdot 2i\epsilon^{\dagger}\theta + \sqrt{2}\epsilon F \,.
\end{align}
The corresponding Lagrangian is 
\begin{align}\label{eq:L02B}
	\cL_B &= \frac{1}{2}\int\dd^{2}{\theta} \left[ G(A,A^{\dagger})B^{\dagger}B \right]
	\nonumber\\[2mm]
	&= G \left[ i\bpsi_R\d_L\psi_R + i\bpsi_{R}(\Gamma\d_L\varphi)\psi_R  \right]
	- \frac{1}{2}\rie(\bpsi\psi)^2
\end{align}
where $G(A,A^{\dagger})$ is the metric obtained by promoting $\varphi,\bphi$ to $A,A^{\dagger}$, respectively, in \eqref{eq:metric}. Note that we have integrated out the auxiliary $F$ field. One can then see that the combination of \eqref{eq:L02} and \eqref{eq:L02B} leads to Eq. \eqref{eq:L22}. The enhancement of the supersymmetry from {\ntwoo} to {\ntwot} was first demonstrated in \cite{Cui:2010si} for the undeformed $\cp^1$ case.

\subsubsection{Hypercurrent multiplet}\label{sec:hypercurrent}

In the following, we analyze the {\em hypercurrent} multiplet $\cJ_\mu$ (see \cite{Shifman:2006bs,Shifman:1986zi,Komargodski:2010rb,Dumitrescu:2011iu} for review and examples) of the deformed $\cp^1$ model. Our discussion on the case of {\ntwot} will run parallel to that of \cite{Shifman:2006bs}. This supermultiplet contains a $R$-current $v_{\mu}$, a supercurrent $s_{\mu\alpha}$, and the energy-momentum tensor $\vartheta_{\mu\nu}$, 
\begin{align}
	\cJ_{\mu} = v_{\mu} + \left[ \theta\gamma^0s_{\mu} + \mbox{H.c.} \right] - 2\bar{\theta}\gamma^{\nu}\theta\, \vartheta_{\mu\nu} + \cdots 
	\,
\end{align}
where the $\gamma$ matrices are defined as 
\begin{align}
	\gamma^0 = \sigma_2 
	\,,\quad 
	\gamma^1 = i\sigma_1 
	\,,\quad 
	\gamma_5 = \sigma_3
	\,.
\end{align} 
Here the Grassmannian coordinate has two complex components $\theta=(\theta^1,\theta^2)$ in contrast to the case of {\ntwoo} which has only one relevant Grassmannian coordinate.
The lowest component $v_{\mu}$ in the hypercurrent is the vector U(1) current. Although classically $v_{\mu}$ is algebraically related to  the axial current discussed in Sec. \ref{sec:thetaterm}) at the quantum level they are different -- the axial current has an anomaly.\footnote{In fact, there are three independently conserved U(1) currents in this model. The first is the axial current presented in Eq. (\ref{eq:axialcurrent}). It is  
generated by the transformation (\ref{eq:axialu1}), conserved at the classical level and acquires a one-loop anomaly in $\partial_\mu J_5^\mu$. The second conserved current $J^\varphi_{\mu}$ is purely bosonic, it is generated by the transformation $\varphi\to e^{i\beta} \varphi$ and $\bar\varphi\to e^{-i\beta} \bar\varphi$. Needless to say, it is anomaly-free. The third is purely fermionic {\em vector} current, also anomaly-free.}
In the spinorial notation, it takes the form
\begin{align}
	\cJ_{\alpha\beta} 
	= \left( \gamma^0\gamma^{\mu} \right)_{\alpha\beta}\cJ_{\mu} 
	= G \bar{D}_{\alpha}\bar{\Phi}D_{\beta}\Phi
\end{align}
where $D_{\alpha}$ and $\bar{D}_{\beta}$ are superderivatives and $\Phi$ and $\bar{\Phi}$ are the chiral superfields with the lowest components $\varphi$ and $\bphi{}$, respectively. At the classical level, the spinorial components $\cJ_{11}$ and $\cJ_{22}$ are conserved, namely,
\begin{align}\label{eq:hypccls}
	\left[ \bar{D}_2\cJ_{11} \right]_{\mbox{\scriptsize classical}} 
	= \left[ \bar{D}_1\cJ_{22} \right]_{\mbox{\scriptsize classical}} = 0
	\,.
\end{align} 
In $\cJ_{\mu}$, only two diagonal components of $\cJ_{\alpha\beta}$ are relevant.

Quantum mechanically, the hypercurrent expressions in \eqref{eq:hypccls} are  anomalous The anomaly is  exhausted by the one-loop effect. For $\cp^1$ the anomaly equations were derived in \cite{Shifman:2006bs}. In our deformed model the anomaly in the right-hand side takes the form
\begin{align}\label{eq:22anom}
	\bar{D}_2\cJ_{11} = \frac{1}{4\pi}\bar{D}_1\left[  
		\frac{1}{2}G\cR \bar{D}_2\bar{\Phi}D_1\Phi
	\right]
	\,,\quad
	\bar{D}_1\cJ_{22} = \frac{1}{4\pi}\bar{D}_2\left[  
		\frac{1}{2}G\cR \bar{D}_1\bar{\Phi}D_2\Phi
	\right]
	\,.
\end{align}
Note the emergence of the scalar curvature $\cR$ on the right-hand side.

The coefficient in \eqref{eq:22anom} can be verified through the scale anomaly of the energy-momentum tensor i.e. the anomaly in $\gamma_{\mu}s^{\mu}_{\alpha}$ \cite{Shifman:2006bs,Losev:2003gs},
\begin{subequations}\label{eq:emanom}
\begin{align}
	\left( \vartheta^{\mu}_{\mu} \right)_{\rm anom} 
	&= \frac{1}{4\pi}G\cR \left(  
		\d_{\mu}\varphi\d^{\mu}\bphi 
		+ i\bpsi\gamma^{\mu}\nabla_\mu\psi
	\right),
	\\[2mm]
	\left( \gamma_{\mu}s^{\mu} \right)_{\rm anom}
	&= \frac{1}{4\pi}G\cR (\d_{\mu}\bphi)\gamma^{\lambda}\psi
	\,.
\end{align}
\end{subequations}
Note that the hypermultiplet $\cJ_{\mu}$ falls in the class of the $R_V$-multiplets in \cite{Dumitrescu:2011iu} since $\d_{\mu}\cJ^{\mu}=0$.
In fact, Eq. \eqref{eq:22anom} can be recast in the standard form of the hypercurrent multiplet proposed in \cite{Komargodski:2010rb,Dumitrescu:2011iu}. Namely,\footnote{For a general ${\cal N}=(2,2)$ $\sigma$  model, the anomalies of the hypercurrent take the form 
\begin{align}
\label{kvsus}
	\chi_{\beta} = \bar{D}_{\beta}\left(  
		-\frac{1}{8\pi}D^{\alpha}\bar{D}_{\alpha}\log\det G_{i\bar{j}}
	\right)
	\,.
\end{align}
This issue was previously discussed in \cite{Komargodski:2010rb} for the case of the symmetric K\"ahler manifolds. Since in our deformed $\cp^1$ model the target space is non-symmetric, the expression in the right-hand side of (\ref{kvsus}) is somewhat different from that in \cite{Komargodski:2010rb}.}
\begin{align}\label{eq:s22anom}
	\bar{D}^{\alpha}\cJ_{\beta\alpha} = \chi_{\beta}
	\qq{with}
	\chi_{\beta} = \bar{D}_{\beta}\left(  
		-\frac{1}{4\pi}D^{\alpha}\bar{D}_{\alpha}\log G
	\right)
\end{align}
for which we use the fact that only the twisted chiral (antichiral) part of $D^{\alpha}\bar{D}_{\alpha}$ contributes in the first (second) equations in \eqref{eq:22anom}.

\begin{center}
*****
\end{center}

The hypercurrent for the $\cp^1$ ${\cal N}=(0,2)$ model was discussed in \cite{Cui:2011uw,Chen:2014efa}. Taking into account our Lie-algebraic extension we arrive at the classical expressions
\begin{subequations}\label{eq:02hypc}
\begin{align}
	\cJ_2 
	&= \frac{1}{2} \eval{\cJ_{22}}_{\theta^1=0}
	= \frac{1}{2} G \bar{D} A^{\dagger}DA\,,
	\\[2mm]
	\tilde{\cT}_{1111} &= -\frac{1}{2} [\bar{D}_1,D_1]\eval{\cJ_{11}}_{\theta^1=0}
	= G\d_R A^{\dagger}\d_R A\,,
\end{align}
\end{subequations}
where $\cJ_2$ and $\tilde{\cT}_{1111}$ stand for two components in the hypercurrent in the ${\cal N}=(0,2)$ model, and $A$ is the {\ntwoo} superfield defined in Sect. \ref{212}. In the {\ntwoo} superspace, the reduced superderivatives are 
\begin{align}
	D = \pdv{\theta} - i\theta^{\dagger}\d_L 
	\,,\qquad
	\bar{D} = -\pdv{\theta^{\dagger}} + i \theta\d_L
\end{align}
Here the lowest component of $\cJ_2$ is the chiral $U(1)$ current $G\psi^{\dagger}_L\psi_L$ and is \emph{not} conserved as the quantum corrections are taken into account. Also, the bosonic component of $\tilde{\cT}_{1111}$ is the part of the energy-momoentum tensor, $T_{1111}$.

In general, like in the {\ntwot} case, the hypercurrent \eqref{eq:02hypc} is conserved classically and becomes anomalous due to one-loop corrections. In other words, the general anomaly equations turn out to be 
\begin{align}\label{eq:02anom}
\begin{aligned}
	&\d_{R}\cJ_2 = -\frac{1}{2}D_2 X + \frac{1}{2}\bar{D}_2\bar{X}
	\,,\qquad
	\bar{D}_2\tilde{\cT}_{1111} = \d_R X\,,
	\\[2mm]
	& X \equiv -\frac{1}{8\pi}G\cR (\d_R A)\bar{D}A^{\dagger}
\end{aligned}
\end{align}
where $X$ encodes the anomalous part of two real supermultiplets $\cJ_2$ and $\tilde{\cT}_{1111}$. Note that the coefficient of Eq. \eqref{eq:02anom} can be fixed by the anomalous chiral $U(1)$ current $G\psi^{\dagger}_L\psi_L$ in parallel to the consideration of the axial $U(1)$ current in Appendix \ref{sec:acrt}, 
\begin{align}
	\d_R(G\psi^{\dagger}_L\psi_L)
	= 2 \cdot \left( \frac{i}{8\pi} G\cR\epsilon^{\mu\nu}\d_{\mu}\varphi\d_{\nu}\bphi \right)
\end{align}
where the prefactor $2$ indicates the number of the fermion zero modes in the instanton background, which is half of the number in {\ntwot} theory, see also \eqref{eq:j5dq}.

\vspace{2mm}

{\sl Important warning}: in the ${\mathcal N} =(0,2)$ case, the $\beta$ function is not exhausted by one loop, see Sect. \ref{sec:betafunc}. Therefore, the one-loop anomaly expression given in \eqref{eq:02anom} should be understood as an operator expressions subject to further infrared multi-loop corrections, just in the same way as in ${\mathcal N}=1$ super-Yang-Mills (see \cite{SV} and Secs. 10.16.1-10.16.4 in \cite{shU}). In our problem, the latter conjecture is not yet proven.

\subsection{Adding twisted masses}
\label{sec:twistedmass}

% \subsubsection{Twisted mass}

As observed in \cite{Alvarez-Gaume:1983uye,Gates:1983py,Gates:1984nk,Dorey:1998yh}, one can introduce the twisted mass parameter consistent with the underlying supersymmetries in the presence of the $U(1)$ isometry of the system. That is, the infinitesimal transformations read 
\begin{align}
	\delta\varphi = it_1\varphi
	\,,\quad
	\delta\bphi = -it_1\bphi
\end{align}
where $t_1$ is the variable parametrizing the isometry. Notice that for the deformed $\cp^1$ model, such an isometry can be summarized by the Killing potential $D(\varphi,\bphi)$,
\begin{align}\label{eq:momentmap}
    D(\varphi,\bphi) &= 
    \frac{1}{2\sqrt{k^2-1}} 
    \log(\frac{\abs{\varphi}^2+k-\sqrt{k^2-1}}{\abs{\varphi}^2+k+\sqrt{k^2-1}})
\end{align}
generating the Killing vectors, namely,
\begin{align}
	\dv{\varphi}{t_1} = -iG^{-1}\pdv{D}{\bphi{}}
    \,,\quad
    \dv{\bphi{}}{t_1} = -iG^{-1}\pdv{D}{\varphi}
	\,.
\end{align}
Here the Killing potential is defined up to an additive constant.

One can then introduce a constant auxiliary vector multiplet $V$ parametrized by the twisted masses, $m$ and $\overline{m}$, to modify the $U(1)$ invariant combination $\abs{\Phi}^2$ in the associated K\"ahler potential as $\Phi^{\dagger}e^{V}\Phi$. In the following, we directly quote the resulting Lagrangian. The interested readers can find a concise review in Sec. 2 of \cite{Shifman:2006bs}. The deformed $\cp^1$ model with the twisted masses is formulated as follows\,\footnote{Generally speaking, for a sigma model with a complex target space with the Killing vectors $X^i,\bar{X}^{\bar{j}}$ for the $U(1)$ isometries the associated Lagrangian takes the form 
\begin{align*}
	\cL_{m} = G_{i\bar{j}}\left[ 
		\d_{\mu}\varphi^{i}\d^{\mu}\bphi^{\bar{j}} 
		- \abs{m}^2X^{i}\bar{X}^{\bar{j}}
		+ i\bpsi^{\bar{j}}\cancel{\nabla}\psi^i 
		-i (D_{k}X^i)\bpsi^{\bar{j}}\psi^k
	\right]
	- \frac{1}{2}R_{i\bar{j}k\bar{l}}\bpsi^{\bar{j}}\psi^{i}\bpsi^{\bar{l}}\psi^{k}
\end{align*}
where $D_kX^i = \d_kX^i + \Gamma^{i}_{kj}X^j$ is the covariant derivative on the target space.}

\begin{align}
\label{eq:lametm}
    \cL_{m} 
    &= 
        G \left[ \d_{\mu}\varphi\d^{\mu}\bphi 
        - \abs{m}^2\varphi\bphi{}
        + i\bpsi\cancel{\nabla}\psi 
        - \left( 1+\Gamma\varphi \right)\bpsi\tilde{\mu}\psi
        \right] 
        - \frac{1}{2}\rie\left( \bpsi\psi \right)^2
\end{align}
where $\nabla_{\mu}$ is the covariant derivative
\begin{align}
	\nabla_{\mu}\psi = \d_{\mu}\psi + (\Gamma\d_{\mu}\varphi)\psi 
\end{align}
and 
\begin{align}
	\tilde{\mu} \equiv m \frac{1-\gamma_5}{2} + \overline{m}\frac{1+\gamma_5}{2}
	\,.
\end{align}

Similarly, one can introduce the twisted mass for the {\ntwoo} model by replacing the aforementioned constant auxiliary {\ntwot} vector multiplet with a {\ntwoo} one. As a result, the {\ntwoo} model with twisted mass takes the form 
\begin{align}
	\cL_{m,(0,2)} = G \left[ \d_{\mu}\varphi\d^{\mu}\bphi 
	- m^2\varphi\bphi{}
	+ i\bpsi_L\nabla_R\psi_L
	- m\left( 1+\Gamma\varphi \right)\bpsi_L\psi_L
	\right] 
	\,
\end{align}
where $m$ is real in the {\ntwoo} case.

\subsection{The \boldmath$\theta$ term}
\label{sec:thetaterm}

The $\theta$ term can be added in  a straightforward manner \cite{Shifman:2006bs,Losev:2003gs},
\begin{align}\label{eq:thetaterm}
	\cL_{\theta} &= \frac{i\theta}{8\pi}G\cR\dd{\varphi} \wedge \dd{\bphi}
	\,.
\end{align}
Note that the theta term is topological and invariant under $2\pi\ZZ$ translation since the topological charge defined as 
\begin{align}\label{eq:topoc}
	Q \equiv \frac{1}{8\pi} \int G\cR \dd^2{\varphi} 
	\in \ZZ
	\,.
\end{align}
In particular, $\abs{Q} = 1$ for the (anti-)instanton solution.
% \begin{align}\label{eq:otopoc}
% 	\frac{1}{8\pi} \int G\cR \dd^2{\varphi} = 1
% 	\,
% \end{align}
% for a single instanton solution. Thus, we can define the topological charge of the system as
For completeness, the theta term can also be expressed as a total derivative
\begin{align}
	R_{1\bar{1}}\dd{\varphi} \wedge \dd{\bphi} 
	= \dd \left(
		- \frac{2n_1+n_2\abs{\varphi}^2}{n_1+n_2\abs{\varphi}^2+n_3\abs{\varphi}^4}
		\cdot \dd{\log{\varphi}}
	\right)
	\,.
\end{align} 
This also indicates that \eqref{eq:topoc} saturates at the small field configuration.

The theta term \eqref{eq:thetaterm} can be further utilized to detect the number of fermion zero modes and then the number of the bosonic zero modes via supersymmetry. To this end, let us consider the divergence of the axial $U(1)$ current 
\begin{align}\label{eq:axialcurrent}
	J_{5}^{\mu} \equiv G\bpsi\gamma^{\mu}\gamma_5\psi
\end{align}
generated by the $U(1)_A$ transformation 
\begin{align}\label{eq:axialu1}
	\psi \to e^{i\alpha\gamma_5}\,\psi 
	\,,\quad
	\bpsi \to \bpsi\, e^{i\alpha\gamma_5}
\end{align}
where $\alpha$ is the variable parametrizing the transformation.
Classically, $j_5^\mu$ is conserved, but it becomes anomalous as the quantum effects are taken into account.  Namely, in {\ntwot}, 
\begin{align}\label{eq:j5dq}
	\d_{\mu}J_5^\mu = 4 \cdot \left( 
		-\frac{i}{8\pi} G\cR \epsilon^{\mu\nu}\d_{\mu}\varphi\d_{\nu}\bphi
	\right)
\end{align}
implying the number of the fermion zero modes is four and therefore the same for the bosonic sector through supersymmetry. See Appendix \ref{sec:acrt} for further detail. Comparing with Eq. \eqref{eq:thetaterm}, one sees that $\theta \to \theta - 4\alpha$ under the axial $U(1)_A$ rotation \eqref{eq:axialu1}, which breaks $U(1)_A$ into $\ZZ_4$.

Furthermore, it is worth pointing out how the instanton action is related to the topological charge defined in \eqref{eq:topoc}. According to the standard BPS argument, the action satisfies 
\begin{align}\label{eq:BPS}
	S &=  \int G 
	\left[ 
		\frac{1}{2}\abs{\d_{\mu}\varphi \pm \epsilon_{\mu\nu}\d^{\nu}\varphi}^2
		\mp \epsilon_{\mu\nu}\d^{\mu}\varphi\d^{\nu}\bphi{}
	\right]
	\dd^{2}{x}
	\nonumber\\[2mm]
	& \geq \frac{
		2\pi\log\left[ \frac{n_2}{2n_1n_3}\left( n_2 +  \sqrt{n_2^2-4n_1n_3}\right) - 1 \right]
	}{\sqrt{n_2^2-4n_1n_3}}
	\cdot \abs{Q}
\end{align}
where $Q$ is the topological charge. The action saturates the BPS bound for  the instanton configuration. In the non-degenerate case (i.e. $n_1,n_3 \neq \infty, 0$), the overall coefficient in front of $|Q|$ in \eqref{eq:BPS} is 
\begin{align}
	\frac{4\pi\arccosh{k}}{g^2\sqrt{k^2-1}}
	\qfor 
	k \geq 1
\end{align}
as was previously derived in \cite{g3}. The $\cp^1$ expression can be obtained by further taking the $k \to 1$ limit.

\section{Analysis of the two-loop beta function}
\label{sec:betafunc}

\subsection{The two-loop beta function of the bosonic model from supersymmetry}
\label{sec:betafunc1}

We will start from the universal fact that the second (and all higher) coefficients of two-dimensional ${\cal N}= (2,2)$ sigma models vanish 
\cite{Alvarez-Gaume:1981exa,Friedan:1980jm}, see also \cite{Bykov:2023klm,Alfimov:2023evq} for recent discussions.\footnote{
In the mid-1980s this fact was questioned by Grisaru, van de Ven and Zanon \cite{Grisaru:1986px} who analyzed the $\beta$ functions in two-dimensional
K\"ahler $\sigma$ models up to four loops. In the case of the  Ricci-flat manifolds they arrived at the conclusion that there is a non-vanishing contribution to the $\beta$ function
cubic in the Riemann curvature of the target space
at the {\em fourth} loop. This result is in the direct contradiction with those reported in \cite{AG1,AG2}. The class of models we study is {\em not} Ricci-flat.
Morever, in Sec. \ref{comment} we will argue that the result \cite{Grisaru:1986px} cannot be applied to the model under discussion. Our argument is based on  the Dorey's exact solution \cite{Dorey:1998yh} for the mass spectrum in $\cp^{N-1}$ \`a la Seiberg-Witten type \cite{sw1,sw2}.
} 
In supersymmetric 
sigma models the two-loop contributions to the $\beta$ function can be  separated into the (purely) bosonic $\beta_{2,b}$ and fermionic $\beta_{2,f}$ parts. In other words, 
\begin{align}
	\beta^{(2)} = \beta_{b}^{(2)} + \beta_{f}^{(2)}
	\overset{!}{=} 0 
\end{align}
which implies, in turn, that the purely bosonic component can be extracted from the fermionic part (which is much more amenable for loop calculations),
\begin{align}\label{eq:bf}
	\beta_{b}^{(2)} = - \beta_{f}^{(2)} \,.
\end{align}
Note that $\beta_b^{(2)}$ is identical to the beta function of the non-supersymmetric case.

\begin{figure}[t] 
  \centering 
  \includegraphics[width=.4\linewidth]{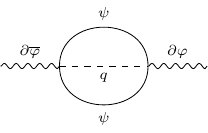}
  \caption{\small The two-loop fermion diagram in a background field calculation.}
  \label{fig:twoloopf}
\end{figure}

To confirm the previous assertion, let us apply the background field method. It suffices to consider the two-loop fermionic diagram in Fig. \ref{fig:twoloopf} -- the only non-trivial diagram with the requred logarithmic divergence.
{\noindent}Here $\d\varphi,\d\bphi$ are the external legs, and $q$, $\psi$, and their complex conjugates are the quantum scalar field and fermions, respectively. To proceed, we consider the background expansion via the K\"ahler coordinates \cite{Higashijima:2000wz,Higashijima:2002fq}. The only relevant interactions in Fig. \ref{fig:twoloopf} are the 3-vertices, namely,
\begin{align}
	i\rie \left( \bq\d_{\mu}\varphi - q\d_{\mu}\bphi \right)\left( \bpsi{}\gamma^{\mu}\psi \right)
	\label{33}
\end{align}
where $\rie = \rie(\varphi,\bphi)$. Therefore, the two-loop fermion correction to the Lagrangian \eqref{eq:L22} is 
\begin{align}
	i\Delta S_{2,f} &= \int\dd^{2}{x}\dd^{2}{y} \,(\rie)^2\d_{\mu}\varphi\d_{\nu}\bphi{} \expval{
        \left( \bq\bpsi{}\gamma^{\mu}\psi \right)_x
        \left( q\bpsi{}\gamma^{\mu}\psi \right)_y
    }
    \nonumber\\[2mm]
	&= -\frac{i}{8\pi^2\epsilon}\int\dd^{2}{x}(\rie)^2G^{-3}\d_{\mu}\varphi\d^{\nu}\bphi{}
	+ \cdots
\end{align}
where the ellipses stand for non-logarithmic divergences and $\epsilon = 2-D$ in dimensional regularization. In accordance with the renormalization group equation (see e.g. \cite{SK,Alvarez-Gaume:1981exa}), we then have 
\begin{align}\label{eq:bf2}
    \beta_{f}^{(2)} = 2 \cdot \left( -\frac{1}{8\pi^2} \right) (\rie)^2G^{-3}
    = -\frac{1}{16\pi^2}G\cR^2
	\,.
\end{align} 
Note that to get the second equality, the first equation in \eqref{eq:rie} is used.

Now, invoking (\ref{eq:bf}) we arrive at
\begin{align}
	\beta_{b}^{(2)} = -\beta_{f}^{(2)} = \frac{1}{16\pi^2}G\cR^2 \,
	\label{36}
\end{align}
which matches with the general formula in the purely bosonic model \cite{SK},
\begin{align}
	\beta_{b}^{(2)} = -\frac{1}{4\pi^2}R_{1\bar{\mu}\nu\bar{\lambda}}R_{\bar{1}}^{~\bar{\mu}\nu\bar{\lambda}}
	= \frac{1}{16\pi^2}G\cR^2 \,.
	\label{37}
\end{align}
In the present case, all the Greek indices $\mu,\nu,\lambda$ are 1.
As a consistent check, taking the limit $k=1$ 
in (\ref{5five}) we recover the $\cp^1$ result\,\footnote{This calculation for $\beta_{b}^{(2)}$ in $\cp^1$ was first presented in \cite{shU}, page 265, see also subsection \ref{vao}.} 
\begin{align}\label{eq:cp2bf}
	\beta^{(2)}\left( \frac{2}{g^2} \right) = \frac{g^2}{2\pi^2}
	\implies 
	\beta^{(2)}(g^2) = -\frac{g^6}{4\pi^2}
	\,.
\end{align}

As explained in detail in \cite{shU} on pages 674-676, the calculation of the graph in Fig. \ref{fig:twoloopf} becomes transparent if we first deal with the fermion loop keeping fixed the momentum flowing through the dashed (bosonic) line. As is clear from Eq. (\ref{33}) the fermion loop is exactly the same as in the two-dimensional Schwinger model. It has no logarithms and is saturated in the infrared. Including the bosonic loop provides us with the first power of $\log\mu$.
This is exactly what is expected in the two-loop graph for the $beta$ function. The Schwinger ``anomaly'' is crucial.

\subsubsection{Verification around the origin}
\label{vao}

If we limit ourselves to the vicinity of the origin in the target space and forget for a short while about the target space invariance, the proof of our assertion can be greatly simplified. Indeed, because the overall structure of the field dependence is constrained by the target space geometry of the deformed $\cp^1$ model, we can {\em accept}  that the second coefficient of the beta function in the bosonic model takes the form 
\begin{align}
	\beta^{(2)} = c_2G\cR^2 \,.
\end{align}
Plugging the explicit expression for the  geometric data given in Eq. \eqref{eq:23}, one sees that 
\begin{align}
	\beta^{(2)} \to c_2 \cdot \frac{4n_2^2}{n_1}
\end{align} 
for $\varphi,\bphi \approx 0$. 
In  the same approximation, the leading terms in the Lagrangian are 
\begin{align}
	\cL_{(2,2)} &= 
	\frac{1}{n_1} \left( \d_{\mu}\varphi\d^{\mu}\bphi + i\bpsi\cancel{\d}\psi \right)
	- i\left( \frac{n_2}{n_1^2} \right) \bphi\d_{\mu}\varphi \left( \bpsi\gamma^{\mu}\psi \right)
	+ \cdots
\end{align}
Then, considering the same two-loop diagram in Fig. \ref{fig:twoloopf}, we obtain the two-loop Lagrangian   
\begin{align}
	\Delta\cL &= \left[  -2 \cdot \left( \frac{n_2}{n_1^2} \right)^2 T\left\{ \left( \bphi\bpsi\gamma^{\mu}\psi \right),\left( \varphi\bpsi\gamma^{\nu}\psi \right) \right\} \right] \d_{\mu}\varphi\d_{\nu}\bphi
	\nonumber\\[2mm]
	&= - \left( \frac{2n_2^2}{n_1} \right)\d_{\mu}\varphi\d^{\mu}\bphi \cdot \frac{1}{8\pi^2}\log\frac{M}{\mu}
\end{align}
where $M$ and $\mu$ are the ultraviolet and infrared cutoffs, respectively. The constant $c_2$ turns out to be 
\begin{align}
	c_2 = -\frac{1}{16\pi^2}
\end{align}
consistent with the covariant derivation given in \eqref{eq:bf2}.

\subsubsection{The two-loop beta function of \boldmath{\ntwoo} extension}
\label{tlbf}

Based on the result of Eq. \eqref{eq:bf2}, the second coefficient of the beta function for minimal {\ntwoo} sigma models with a one-complex-dimensional target space can be readily identified. To see that this is the case, note that the fermion sector in {\ntwot} models consists of two Weyl fermions (one left-handed fermion and one right-handed) while in {\ntwoo} models, there exists only one Weyl fermion. This indicates that the contribution from fermions at the two-loop level is half of Eq. \eqref{eq:bf2} in the {\ntwoo} case. Consequently, combining with the bosonic contribution, we obtain the second coefficient of the beta function that 
\begin{align}
	\beta^{(2)}_{(0,2)} = \left( 1 - \frac{1}{2} \right) \cdot \frac{1}{16\pi^2}G\cR^2
	= \frac{1}{32\pi^2}G\cR^2 \,.
\end{align}
Going through the same process around \eqref{eq:cp2bf}, one would see the second coefficient of the {\ntwoo} $\cp^1$ model
\begin{align}
	\beta^{(2)}_{(0,2)}(\cp^1) = -\frac{g^6}{8\pi^2}
\end{align}
which was first derived in \cite{Cui:2011uw} through the superfield calculation.

\subsection{Comparison with the first-order formalism}

Our results (\ref{36}) for the bosonic model coincides with that obtained in \cite{g2} by virtue of the first-order formalism.
The regularization procedure used in \cite{g2} was as follows. We start from the {\ntwot} theory. In first-order formalism it is obvious that all loops in  $\beta$ beyond the first loop 
vanish -- there is no anomaly and holomorphy is preserved. Then we endow the fermion field with a mass term $m_f$ and compute 
the $\beta$ function coefficients with large but fixed value $m_f$. In the limit $m_f\to\infty$ we discover that some ``extra'' terms do not vanish. It is just these extra terms which are responsible for the transition from $\pt_i v_a^j\pt_jv_b^i$ in (\ref{9nine}) to  $\nabla_i v_a^j\nabla_jv_b^i$ where $\nabla_\ell$ stands for the covariant derivative. This procedure can be viewed as an ultraviolet derivation of the anomaly.

\subsection{Comment on the literature}
\label{comment}

We started Sec. \ref{sec:betafunc1} from the statement that ``the second (and all higher) coefficients of two-dimensional ${\cal N}= (2,2)$ sigma models vanish.'' 
In this subsection we will discuss this statement in more detail. A series of papers on this subject was published in \cite{AG1,AG2,AG3} in the early 1980s. Then this issue was revisited  1985-86, approximately simultaneously with the publication \cite{Grisaru:1986px}. 

The authors of \cite{Grisaru:1986px} state the opposite --  that 
 the $\beta$ functions in two-dimensional {\em Ricci flat}
K\"ahler $\sigma$ models have a non-vanishing contribution at four loops
cubic in the Riemann curvature of the target space, see their Eq. (5.16).

The class of Lie-algebraic models we are interested in (it includes, in particular $\cp^{N-1}$ models) is {\em not} not Ricci-flat.
In 37 years that elapsed since the publication of Grisaru et. al. a significant progress happened in understanding both perturbation theory and {\em exact} solutions
in $\cp^{N-1}$ models (and their extensions) and, in  ``parallel'' to them, exact solutions in Yang-Mills theories with various degrees of supersymmetry.

If supersymmetry is minimal, the $\beta$ functions are indeed multi-loop, but are exactly calculable. For ${\cal N}=2$ supersymmetry the perturbative $\beta$ 
functions are exhausted by the first loop. This is seen from the analysis of the holomorphy properties with regards to the complexified coupling constant
$$1/g^2_{\rm holom} = 1/g^2 +i\theta/(8\pi^2)$$  
  in super-Yang-Mills and
  $$ 1/g^2_{\rm holom} = 1/g^2 +i\theta/(4\pi) $$
in 2D $\cp^{N-1}$ models (see below).  Moreover, this statement is confirmed 
by the exact solutions.

The exact solution for the mass spectrum of the Seiberg-Witten ${\cal N}=2$ super-Yang-Mills \cite{sw1,sw2} (say, for SU(2)$_{\rm gauge})$,
parametrized by a single modular invariant $u$, being expanded in the ratio $u/\Lambda^2$ exhibits
the {\em  first} order in $\log(u/\Lambda)$ plus all powers of 
\beq
(u/\Lambda)^{4n}, \quad n=1,2,3,4, ...
\label{exppt}
\eeq
The power terms of the expansion (\ref{exppt}) do not contain logarithms and come from instantons (this series can be -- and in fact, has been -- obtained by using
the Nekrasov localization \cite{nl}).

Next, in 1998 Dorey published a paper
\cite{Dorey:1998yh} in which he obtained the exact solution \`a la  Seiberg-Witten for the ${\cal N} = (2,2)$ $\cp^{N-1}$ models with twisted masses
(in $\cp^1$ there is only one twisted mass parameter).
Dorey's  method repeats the Seiberg-Witten's analysis \cite{sw1,sw2} in ${\cal N}=2$ Yang-Mills step by step. 

If one replaces the twisted mass of $\cp^{N-1}$ model by the modular parameters $u_i $ of the Seiberg-Witten derivation in Yang-Mills, then the formula for the spectrum in 
Yang-Mills in four dimensions is exactly the same as the Dorey's formulae in $\cp^{N-1}$ in two dimensions (see Eqs. (112) in the general case and (117) for a particular case of $\cp^1$ in \cite{Dorey:1998yh}.)

Dorey's observation  \cite{Dorey:1998yh} can be summarized as follows:

The mass spectrum on the Coulomb branch ($\xi=0$ where $\xi$ is the Fayet-Iliopoulos term) of the Seiberg-Witten theory, with unconfined `t Hooft-Polyakov-like monopoles and dyons coincides with that of $\cp^{N-1}$ models
 emerging on the vortex string \cite{ShiY,HT}  in the Higgs phase (i.e. $\xi\neq 0$). In \cite{ShiY} it was proved that the central charges cannot depend on the non-holomorphic parameter $\xi$ 
 in the BPS sector. 
 This established a one-to-one correspondence between the mass spectra of the two  seemingly different theories.
They prove to be identical in the BPS sectors, hence,  $2D$-$4D$ correspondence. 
 
 Dorey's formula for the BPS mass spectrum 
depends on the ratio $m/\Lambda$, where $m$ is the twisted mass, and $\Lambda$ is the scale parameter of the theory, obtained through the dimensional transmutation (as in any asymptotically free theory). 
Let us assume that this parameter is large and expand the Dorey's exact solution  in the ratio $m/\Lambda$. 
In parallel with ${\cal N}=2$ Yang-Mills the Dorey expansion contains the {\em  first} order in $\log(m/\Lambda)$ plus powers of 
$$
(m/\Lambda)^{2nN}, \quad n=1,2,3,4, ...
$$
where $N$ comes from $\cp^{N-1}$, with nothing else. 

We emphasize, that
the perturbative term is $\log(m/\Lambda)$ is unique. There are no terms $\log^2$ or $\log\log$, etc. in the expansion of the exact formula.
Since the masses are physically observable, their expression must be consistent 
with the $\beta$ function. This can only happen if the perturbative $\beta$ function is purely one-loop in the class of models under consideration.

Just for completeness, let us mention that in minimal supersymmetries (such as ${\cal N}=1$ Yang-Mills or ${\cal N}=(0,2)$ $\cp^1$)
the $\beta$ functions contain all loops. However, if it were not for holomorphic anomaly \cite{ha1,ha2}, all coefficients, starting from the two-loop coefficient would vanish --
only the one-loop coefficient would survive. The break down of holomorphy is an infrared effect \cite{ha1,ha2}, which is well understood. This is best 
illustrated by the instanton formula (IR is automatically regularized in the instantion background). Its 
general form valid both for super-Yang-Mills and $\cp^{N-1}$ (without matter fields) is as follows,
 \beq
  \beta (\alpha) = - \left(n_b-\frac{n_f}{2}\right)\, \frac{ 
\alpha^2}{2\pi}\left[
1-\frac{\left(n_b-n_f\right)\,\alpha}{4\pi}\right]^{-1}\, ,
\label{totbetapg3}
\eeq
where $n_b$ and $n_f$ are the numbers of the bosonic and fermionic zero modes, respectively. 
Above, $\alpha=g^2/(4\pi)$ in  super-Yang-Mills and  $\alpha=g^2/2$ in $\cp^{N-1}$.

All coefficients in the $  \beta$ function (\ref{totbetapg3}) are {\em integers} and, moreover,  of a purely geometric nature. They are in one-to-one correspondence with the number of symmetries nontrivially realized on the BPST or BP instanton. Equation (\ref{totbetapg3}) is valid for
${\cal N} = 2$ and $4$ ($n_b=n_f$ and $n_b=\frac 12 n_f$, respectively). 

For the minimal supersymmetriy (${\cal N}=1$ in $4D$ Yang-Mills) it stays valid too and presents an all-loop $\beta$ function in the form of the geometric progression \cite{nsvzp,st1,st2,st3}. Minimal supersymmetry in the class of $\sigma$ models is ${\cal N}=(0,2)$.
Only $\cp^1$  and its generalizations can be considered in this class since
$\cp^{N-1}$ with $N\geq 3$ do not allow the {\em minimal} ${\cal N}=(0,2)$ super-extension  because of the geometric anomalies (see \cite{Chen:2014efa,Chen:2015xda,Chen:2015dti} and references therein). In Ref. \cite{g3} it is demonstrated in detail that
the term in the square brackets in Eq. (\ref{totbetapg3}) remains  intact  in the Lie-algebraic deformation of $\cp^1$ (see Eq. (49) in \cite{g3}).

\section{Reduction to quantum mechanics}
\label{sec:lame}

In this section, we explore quantum mechanics (QM) associated with the deformed $\cp^1$ model, derived through compactification along the spatial dimension. Under a particular scheme of compactifications, the resulting quantum mechanical system is the Lam\'e QM problem which is Lie-algebraic and quasi-exactly solvable. It can also be viewed as the interpolation between two solvable quantum mechanics, the sine-Gordon and the P\"oschl-Teller systems \cite{Muller-Kirsten:2012wla}\,\footnote{The P\"oschl-Teller case, i.e. $k \to \infty$, is \emph{not} periodic.}. For additional insights regarding the connection to other integrable and Lie-algebraic models, interested readers are referred to earlier discussions from the 90s \cite{Brink:1997zi,Turbiner:1992hg,Turbiner:1994gi}.

For the time being, let us consider only the bosonic version of the deformed $\cp^1$ model with non-singular parameters, i.e. $n_1,n_3$ are neither zero nor infinity. Eq. \eqref{3three} can be recast via the field redefinition \cite{g3} such that the Lagrangian reads 
\begin{align}\label{eq:lb}
	\cL_b = \frac{2}{g^2_{2d}}\frac{\d_{\mu}\varphi\,\d^{\mu}\bphi}{1+2k\abs{\varphi}^2+\abs{\varphi}^4}
\end{align} 
where the parameters $n_i$ are 
\begin{align}
	n_1 = n_3 = \frac{g^2_{2d}}{2}
	\qand
	n_2 = g^2_{2d}k
\end{align}
in which $k \in [1,\infty)$.
Then to understand the connection between the deformed $\cp^1$ model and the Lam\'e equation, one can take the following reparametrization of our main model. Namely, 
\begin{align}\label{eq:repara}
	\varphi(t,z) = \frac{\sqrt{1-\kappa}\sd(\theta(t,z)|\kappa)}{1+\cd(\theta(t,z)|\kappa)}e^{i\alpha(t,z)}
	\,,\quad
	\bphi(t,z) = \frac{\sqrt{1-\kappa}\sd(\theta(t,z)|\kappa)}{1+\cd(\theta(t,z)|\kappa)}e^{-i\alpha(t,z)}
\end{align}
where $\sd(\theta|\kappa)$ and $\cd(\theta|\kappa)$ are two kinds of Jacobi elliptic function and $\alpha$ is the azimuthal angle. The parameter $\theta$ is defined on $[0,2K(\kappa))$ where $K(\kappa)$ is the complete elliptic integral of the first kind and the other parameter $\kappa$ is associated with the original elongation factor $k$ in the way 
\begin{align}
	\kappa \equiv \frac{k-1}{k+1} \in [0, 1) \,.
\end{align}
The conventions and further properties of the Jacobi elliptic functions and their integrals used in this paper are summarized in Appendix \ref{sec:jef}. Plugging \eqref{eq:repara} into the Lagrangian \eqref{eq:lb}, we can write down the bosonic Lagrangian in terms of $\theta$ and $\alpha$,
\begin{align}\label{eq:Lta}
	\cL_b = \frac{2}{g^2_{2d}(1+k)} \Bigl[  
		\d_{\mu}\theta\,\d^{\mu}\theta 
		+ \sn^{2}(\theta|\kappa)\d_{\mu}\alpha\,\d^{\mu}\alpha
	\Bigr]
	\,.
\end{align}
As shown in \eqref{eq:Lta}, we already observe some signals of the emergence of Lam\'e potential as the coefficient of the kinetic term of $\alpha$.

Next, we can apply the Scherk-Schwarz dimensional reduction \cite{Scherk:1979zr} such that the underlying spacetime is $\RR \times S^1_{L}$ where $L$ is the circumference of the compactified circle and the spacetime dependences of $\theta$ and $\alpha$ fields are restricted
\begin{align}\label{eq:bc}
	\theta(t,z) = \theta(t)
	\,,\quad
	\alpha(t,z) = \alpha_0 - \alpha_1 z
	\,,
\end{align}
where $\alpha_0$ and $\alpha_1$ are real time-independent constants and the latter one is constraint by the boundary condition along $S^{1}_L$.
For example, the periodic boundary condition on $\varphi,\bphi$ 
\begin{align}
	\alpha(t,z+L) = \alpha_0 - \alpha_1(z+L)
	\overset{!}{=}
	\alpha_0 - \alpha_1 + 2\pi n
	\,,\quad
	n \in \ZZ
\end{align}
implies that 
\begin{align}\label{eq:pbc}
	\alpha_1 = \frac{2\pi n}{L} \,.
\end{align}
Similar arguments can be applied to the anti-periodic boundary condition and the twisted boundary conditions.
Note that only a subset of the field configurations in the deformed $\cp^1$ model aligns with the condition specified in Eq. \eqref{eq:bc}. For instance, mutli-fractional instanton solutions are not compatible with the Scherk-Schwarz scenario while composites with pairs of fractional and anti-fractional instantons satisfies the assumption.\footnote{
	This observation can be straightforwardly deduced from the findings related to the $\cp^1$ model \cite{Misumi:2015dua}. As demonstrated in \cite{g3}, the instanton equation for the deformed $\cp^1$ model is identical to that of the original $\cp^1$ model, leading to equivalent (fractional/anti-fractional) instanton solutions. Consequently, the discussions pertinent to the $\cp^1$ model are equally applicable to the deformed variant.
} 

This phenomenon was initially identified in the comparison of the $\cp^1$ model with the sine-Gordon model \cite{Misumi:2015dua}, demonstrating that not all field configurations from the two-dimensional model are preserved under the Scherk-Schwarz reduction. A more detailed comparison of deformed $\cp^1$ quantum mechanics and Lam\'e quantum mechanics is provided in Appendix \ref{sec:dcpvslame}.

To proceed with the dimensional reduction, we then insert \eqref{eq:bc} into the two-dimensional Lagrangian and integrate over the $z$-direction, which leads to 
\begin{align}\label{eq:1dL}
	\cL_1 = \frac{2L}{g^{2}_{2d}(1+k)}\left[  
		\left( \dv{\theta}{t} \right)^2
		-\alpha_1^2\sn^2(\theta|\kappa)
	\right]
	\,.
\end{align}
For what follows it is convenient to denote the one-dimensnional coupling constant
\begin{align}
	\frac{1}{g^2} \equiv \frac{2L}{g^{2}_{2d}(1+k)}
\end{align} 
and rescale the time variable $t \to 2t/g^2$. The system of Eq. \eqref{eq:1dL} can be quantized adhering to the standard lore of the quantum mechanics, wherein the time-independent Schr\"odinger equation is expressed as
\begin{align}\label{eq:lameH}
	\frac{1}{g^2}\left[ -\dv[2]{\theta} + \alpha_1^2\sn^2(\theta|\kappa) \right]\Phi(\theta)
	= E\Phi(\theta)
\end{align}
in which $\Phi(\theta)$ is the corresponding wave-function as a function of the compact coordinate $\theta$. Eq. \eqref{eq:lameH} is recognized in the literature as the Lam\'e model \cite{turb,dunne,Turbiner:1988by,Ganguly:2000qy}. 

\subsubsection*{Two limits of the Lam\'e equation}

\begin{figure}[t] 
  \centering 
  \includegraphics[width=.8\linewidth]{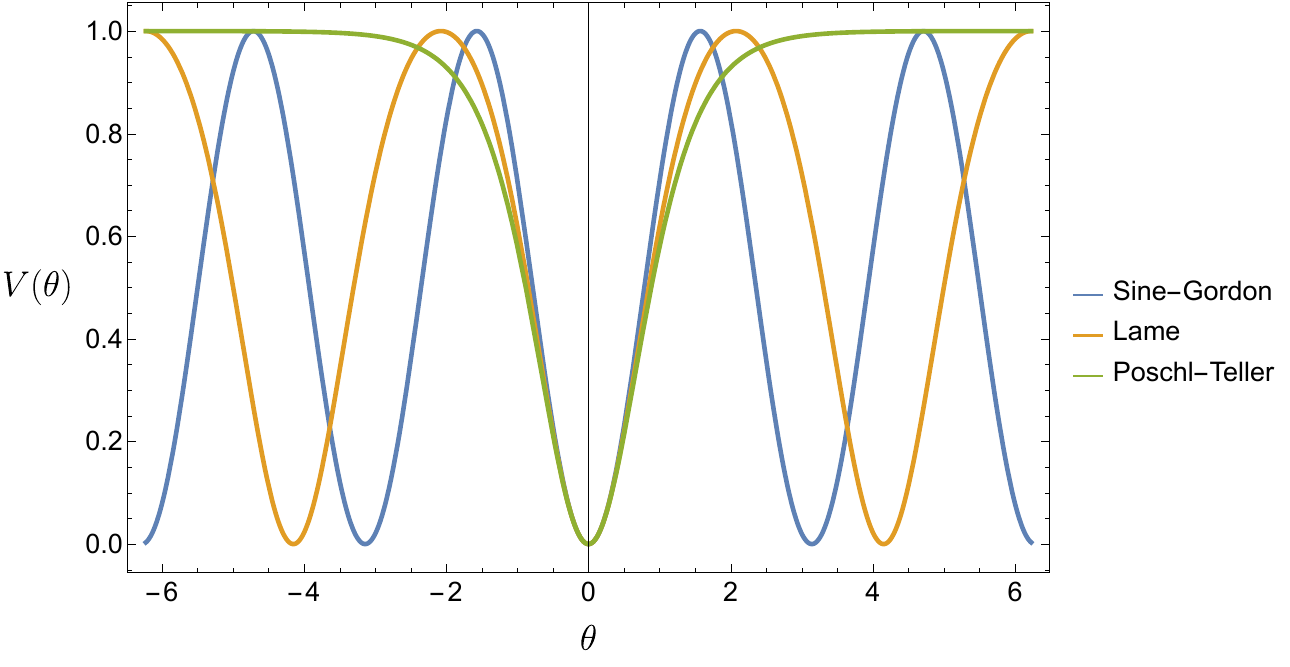}
  \caption{\small The demonstration of the potential of three quantum mechanical systems, sine-Gordon, Lam\'e, and P\"oschl-Teller one. The elliptic modulus of the Lam\'e potential $\kappa$ is 0.7. The blue, orange, and green curves represent the potentials of sine-Gordon, Lam\'e, and P\"oschl-Teller systems, respectively.}
  \label{fig:QMpotntials}
\end{figure}

In the ordinary construction \cite{g1,g2,g3}, we have seen that the two-dimensional deformed $\cp^1$ model is a (Lie-algebraic) generalization of the classic $\cp^1$ model. This can also be realized in its one-dimensional reduction \eqref{eq:lameH}. The figure \ref{fig:QMpotntials} shows the transition of the potentials from the sine-Gordon model through to the Lam\'e one and finally to the P\"oschl-Teller system.

Starting with the limit $\kappa$ approaching zero, one has $\sn(\theta|\kappa) \to \sin{\theta}$ and the Hamiltonian in this case
\begin{align}
	g^2H_{\kappa=0} = -\dv[2]{\theta} + \alpha_1^2\sin^2{\theta}
\end{align} 
in which $\theta \in [0,\pi)$. This Hamiltonian is precisely the one of sine-Gordon quantum mechanics whose potential is periodic. 

On the other hand, the one-dimensional model is also non-trivial in the other limit $\kappa$ reaching the unity. Before proceeding to the reduced quantum mechanics, we briefly review some basic results of the deformed model in the large $k$ limit. From the two-dimensional perspective, the deformed $\cp^1$ model turns out to be the sausage/cigar model \cite{FOZ,LZ} as the elongation $k$ becomes large \cite{g3}. And in the exact limit $\kappa \to 1$, or equivalently $k \to \infty$, the target space of the complex fields $\varphi,\bphi$ deforms to a cylinder \cite{Bykov:2023klm} in the present limit. Namely,
\begin{align}\label{eq:cylinder}
	\cL \sim \frac{\d_{\mu}\varphi\,\d^{\mu}\bphi}{\abs{\varphi}^2} 
	~
	\xleftrightarrow{~~~\varphi = e^{u}~~~}
	~
	\d_{\mu}u\,\d^{\mu}\overline{u}
	\,.
\end{align}
On the one dimensional side, the Schr\"odinger equation \eqref{eq:lameH} can be written as\footnote{We have used the fact that ${\displaystyle \lim_{\kappa \to 1}\sn(\theta|\kappa) = \tanh{\theta} }$.}
\begin{align}\label{eq:ptqm}
	\frac{1}{g^2}\left[ -\dv[2]{x} + \alpha_1^2(1-\sech^2{x}) \right]\Phi(x)
	= E\Phi(x)
\end{align}
in which $x$ now is defined on the half-real line $\RR_{\geq 0}$.
Certainly, the \eqref{eq:ptqm} is reflection invariant, and we can extend the domain from $x \in \RR_{\geq 0}$ to $x \in \RR$. Note that Eq. \eqref{eq:ptqm} is recognized as the P\"oschl-Teller system \cite{Barut:1987am} and is also quasi-exactly solvable for judiciously chosen coefficient $\alpha_1$ (cf. \cite{turb,shifm}.)

\subsection{Comparison to Dunne-Shifman}

As well-known in the literature, the Lam\'e model is Lie-algebraic and quasi-exactly solvable \cite{shifm,turb,Turbiner:1988by,Ganguly:2000qy,dunne}. In other words, the associated Hamiltonian can be expressed as a matrix-valued function of a certain Lie algebra in some representation and a subset of the spectrum can be solved exactly. In \cite{dunne}, the Lam\'e system is studied in detail and shown that there exists a duality between bands and gaps in the spectrum. In the following, we detail the condition when our system is algebraic and its connection to the boundary condition imposed in the Scherk-Schwarz reduction. Without loss of generality, we may set $g^2$ and the circumference of the compactified dimension $L$ to be the unity for convenience.

To this end, recall that in the original construction of the Lie-algebraic sigma model \cite{g1,g2,g3}, the differential representation of $sl(2)$-algebra in the spin-$j$ representation is 
\begin{align}\label{eq:diffrepn}
	T^{+} = 2j\eta - \eta^2\d_{\eta}
    \,,\quad
    T^{0} = -j\eta + \eta\d_{\eta}
    \,,\quad
    T^- = \d_\eta
\end{align}
where $j$ is a semi-integer and 
\begin{align}\label{eq:etavar}
	\eta \equiv 1 - \sn^2(\theta|\kappa)
\end{align}
following \cite{dunne}. Then \eqref{eq:lameH} is recast in the form 
\begin{multline}\label{eq:lamealg}
	H = 4\left[  
        (-1+\kappa)T^0T^- + (-1+2\kappa)T^+T^- - \kappa T^+T^0
    \right]
    \\[2mm]
	+2\left[  
        -(1+6j)\kappa T^+ -2(1+2j)(-1+2\kappa)T^0 + (1+2j)(-1+\kappa)T^-
    \right]
    \\[2mm]
    -4j(1+2j)(-1+2\kappa)+\alpha_1^2
    + \eta \left[  
        4j (4j+1)\kappa - \alpha_1^2
    \right]
	\,.
\end{multline}
For \eqref{eq:lamealg} to be Lie-algebraic, there should be no dependence on the variable $\eta$ in the Hamiltonian. Hence, one requires that\footnote{The dimension of $\alpha_1$ can be recovered by taking $\alpha_1 \to \alpha_1/g^2$.}
\begin{align}\label{eq:alphaj}
	\alpha_1^2 =  4j (4j+1)\kappa 
\end{align}
Note that with the condition \eqref{eq:alphaj}, the coefficient of the potential term in \eqref{eq:lameH} is a multiple of $\kappa$ satisfying the quasi-exactly solvable condition \cite{dunne,shifm}. As discussed in the dimensional reduction process, the constant $\alpha_1$ depends on the boundary condition. Unlike \eqref{eq:pbc} obtained under periodic boundary condition, the condition specified in Eq. \eqref{eq:alphaj} necessitates the imposition of the twisted boundary condition
\begin{align}\label{eq:tbc}
	\varphi(t,z+L) = e^{\pm i \sqrt{4j(4j+1)\kappa}}\varphi(t,z)
	\,.
\end{align}

\subsection{Generalizations}

So far, we have provided a comprehensive discussion on the relation between quasi-exactly solvable quantum mechanics and the Lam\'e quantum mechanics, as derived from dimensional reduction through a specific scheme. In Sec. \ref{sec:extension}, we also see some generalizations to the bosonic deformed $\cp^1$ model. Let us examine how these additional elements in the extended deformed model influence the corresponding quantum mechanics.

\subsubsection{Reduction of the deformed $\cp^1$ model with twisted masses}

In the case of the deformed model with twisted mass \eqref{eq:lametm}, one has 
\begin{align}
	\cL_{m,b} = G\Bigl[
		\d_{\mu}\varphi\d^{\mu}\bphi - \abs{m}^2\varphi\bphi
	\Bigr]
\end{align}
in which the fermionic part is ignored for the time being. Then, taking the same elliptic parametrization \eqref{eq:repara} and going through the same dimensnional reduction process, one deduces the Hamiltonian with twisted mass
\begin{align}
	H_{m} = 
	% \frac{1}{g^2}\left[ 
		-\dv[2]{\theta} + \left( \alpha_1^2 + \abs{m}^2 \right)\sn^2(\theta|\kappa) 
	% \right]
	\,.
\end{align}
By introducing an additional mass parameter, we can relax the twisted boundary condition as specified in Eq. \eqref{eq:tbc}, and instead, define a quantization condition for the mass parameter to satisfy the Lie-algebraic criterion. For example, let us keep the periodic boundary condition \eqref{eq:pbc} intact. Then the quantization condition for the system to be QES is 
\begin{align}
	\abs*{m}^2 = J(J+1)\kappa - \left( \frac{2\pi n}{L} \right)^2
\end{align}
for $J \in \NN$ and $n \in \ZZ$. Similar treatments can be applied to anti-periodic and other boundary conditions along the compactified dimension.

\subsubsection{Supersymmetric Lam\'e model}

\begin{figure}[t] 
    \centering 
    \begin{subfigure}[b]{.47\linewidth}
      \centering
      \includegraphics[width=0.95\textwidth]{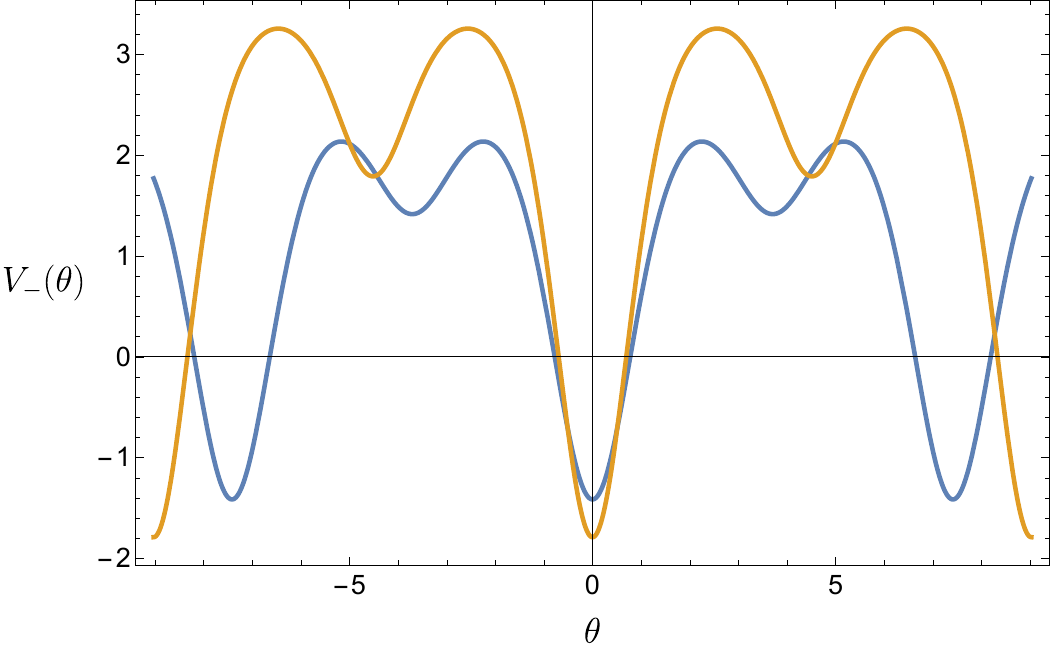}
      \caption{\small $\alpha=2$}
    \end{subfigure}
    \begin{subfigure}[b]{.47\linewidth}
        \centering
        \includegraphics[width=0.95\textwidth]{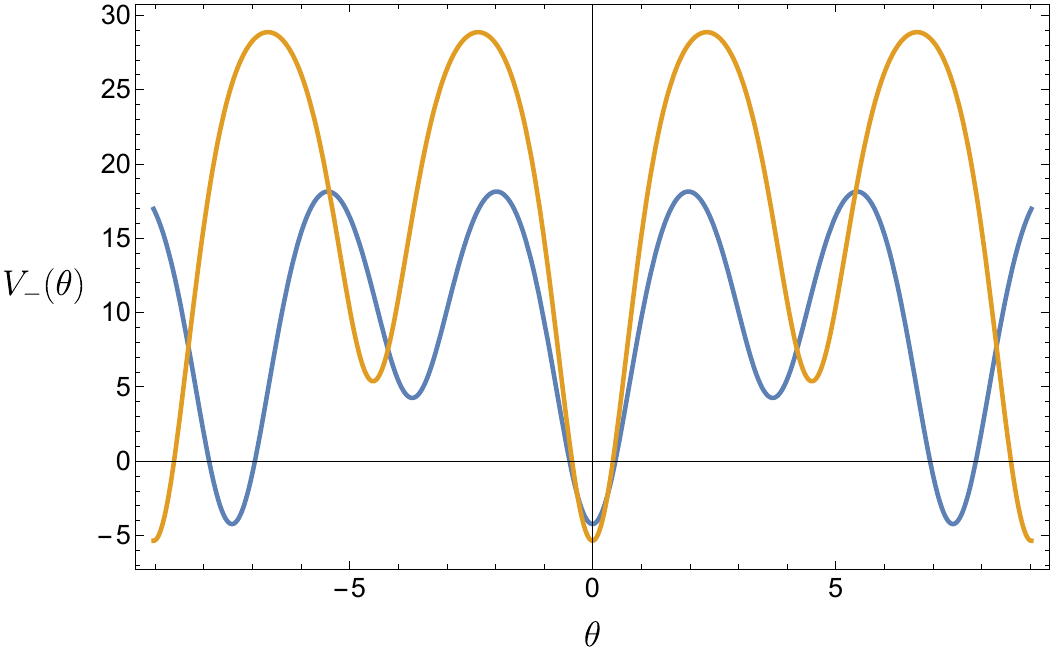}
        \caption{\small $\alpha=6$}
      \end{subfigure}
    \caption{\small The potential $V_-(\theta)$ of the supersymmetric Lam\'e model defined in \eqref{eq:susylame}. The blue and orange lines correspond to the potential with $\kappa$ equal to $0.5$ and $0.8$, respectively. $V_+(\theta)$ has the identical structure to $V_-(\theta)$, but is shifted by a half period.}
    \label{fig:susyqmpot}
\end{figure}

Another direction to generalize the Lam\'e model is to supersymmetrize it. Generally speaking, a supersymmetric quantum mechanics deformed by a potential term takes the form (see for example \cite{hkp})
\begin{align}\label{eq:Ho}
	H_o = -\dv[2]{\theta} + (W'(\theta))^2 + W''(\theta)\begin{pmatrix}
		1 & 0 \\ 0 &-1
	\end{pmatrix}
\end{align}
where $W(\theta)$ is the superpotential and the matrix representation of fermions is adopted. By projecting onto the subspaces of different fermion numbers, the Hamiltonian \eqref{eq:Ho} can be further simplified to an effective bosonic system, namely,
\begin{align}\label{eq:susyqm}
	H = -\dv[2]{\theta} + (W'(\theta))^2 \mp W''(\theta) \,.
\end{align}
In particular, in the supersymmetric Lam\'e quantum mechanics, we have 
\begin{align}\label{eq:susylame}
    H = -\dv[2]{\theta} + V_{\mp}(\theta)
	= -\dv[2]{\theta} + \alpha^2\kappa\sn^2(\theta|\kappa) \mp \alpha\sqrt{\kappa}\cn(\theta|\kappa)\dn(\theta|\kappa)
\end{align}
with the Schr\"odinger equation $H\Phi(\theta) = E\Phi(\theta)$. 
Note that the superpotential\footnote{Here the parameter $\alpha_1 = \alpha\sqrt{\kappa}$ comparing to the previous case.} $W(\theta)$ in our case is 
\begin{align}
    W(\theta) = -\alpha\arctanh(\sqrt{\kappa} \cd(\theta|\kappa)) \,.
\end{align}
With the potential $V_{\pm}(\theta)$ in Eq. \eqref{eq:susylame} at hand, the supersymmetry is manifest, but this is not the case of Lie-algebraicity. The later part of this section is devoted to this issue.

Before moving on to the discussion on the Lie-algebraic structure of the supersymmetric Lam\'e QM, let us have a closer look at other interesting features of this system. First, we note that \eqref{eq:susylame} is compatible with the 2D construction \eqref{eq:L02} and the dimensional reduction scheme. To see this is the case, in addition to \eqref{eq:bc}, one also needs the reduction of the fermions 
\begin{align}\label{eq:fermbc}
	\psi(t,z) = \psi(t)e^{i(\alpha_0-\alpha_1z)}
\end{align}
with the same boundary condition as the scalar fields. As a result, the reduced one-dimensional Lagrangian is 
\begin{align}
	g^{2}\tilde{\cL}_{(0,2)} = 
	% \frac{1}{g^2}\left\{  
		\frac{1}{2}\left[ \dot{\theta}(t) ^2 - \alpha_1^2\sn^2(\theta|\kappa)\right]
        +i\bar{\chi}\dot{\chi}
        -\alpha_1\cn(\theta|\kappa)\dn(\theta|\kappa)\bar{\chi}\chi
	% \right\}
	\,,
\end{align}
where $\chi$ is the normalized fermion field via velbein.
This Lagrangian is obtained from \eqref{eq:L02} by substituting $\varphi$ and $\psi$ within the compactification scheme \eqref{eq:bc} and \eqref{eq:fermbc} and integration over $z$.
It implies that the Hamiltonian indeed matches with the one proposed in \eqref{eq:susylame} by taking $\bar{\chi}$ and $\chi$ to be $\sigma^{+}$ and $\sigma^-$, respectively, in the matrix representation. The ordering of the fermions is fixed such that the original Hamiltonian is $\{Q,\overline{Q}\}/2$ where $Q,\overline{Q}$ are supercharges.

Lastly, the potential of the SUSY quantum mechanical model is depicted in figure \ref{fig:susyqmpot}. It is clear that the potential is periodic with period $4K(\kappa)$ due to it elliptic function nature. As the value of $\alpha$ increases, it is observed that the period is halved compared to cases with smaller $\alpha$. This is because the portion of the potential modified by the supersymmetric effect (or, equivalently, the presence of Weyl fermions) is proportional to $\alpha$, whereas the non-supersymmetric part of the potential is proportional to $\alpha^2$ and has a period of $2K(\kappa)$. Furthermore, if one intend to conduct analysis via the WKB perturbation, it is noteworthy that there exist two distinct types of instanton events in the inverted potential: one involves tunneling through a tall barrier, while the other occurs via a low barrier.

\subsubsection*{Lie-algebraic features of supersymmetric Lam\'e Hamiltonian}
To advance our understanding on the Lie-algebraic nature of the supersymmetric Lam\'e model, let us analyze the potential terms in the Hamiltonian in detail. In accordance with the argument in \cite{Ganguly:2000qy}, the Hamiltonian \eqref{eq:susylame}, especially the last term,  does not match the general form of the quasi-exactly solvable model with a double-periodic potential. However, this is not the end the story. The system can be transformed into a Lie-algebraic form via an appropriate coordinate transformation, though the quasi-exactly solvable condition requires separate verification. The aforementioned assertion is elaborated as follows. Consider the coordinate transformation 
\begin{align}
    \theta' = i\left( \theta - K - iK' \right)
\end{align}
where 
\begin{align}
    K' \equiv K(1-\kappa) = K(\kappa') \,.
\end{align}
Utilizing the identities of Jacobi elliptic functions, we rephrase the corresponding Schr\"odinger equation of \eqref{eq:susylame} in terms of dual variables $\theta',\kappa'$
\begin{align}\label{eq:susylamedual}
    \tilde{H}\Phi = \left[ 
        -\dv[2]{{\theta'}} +  \alpha^2\kappa'\sn^2(\theta'|\kappa')
        \pm i\alpha\kappa'\sn(\theta'|\kappa')\cn(\theta'|\kappa')
    \right]\Phi 
    = \left( \alpha^2 - E \right) \Phi
\end{align}
which fits into the Lie-algebraic form given in \cite{Ganguly:2000qy}. The Hamiltonian \eqref{eq:susylamedual} is not the in the canonical form of SUSY QM. However, we can use the identity of the Jacobi elliptic functions presented in Appendix \ref{sec:jef} such that the dual superpotential is 
\begin{align}
	\widetilde{W}^{\prime}(\theta') = i\dn(\theta'|\kappa') \,.
\end{align}
Then, the Lie-algebraic feature of the system is realized by the similar differential representation in \eqref{eq:diffrepn}, but with a different variable 
\begin{align}
	\xi \equiv \frac{\sn(\theta'|\kappa')}{\cn(\theta'|\kappa')}
	\,.
\end{align}
Note that the Hamiltonian $\tilde{H}$ is formulated in the general form 
\begin{align}\label{eq:tildeH}
	\tilde{H} = -\sum_{a,b=0,\pm} C_{ab}T^aT^b - \sum_{a=0,\pm}C_aT^a - d
\end{align}
with
\begin{align}
\begin{aligned}
	C_{++} &= (1-\kappa')
    \,,\quad
    C_{00} = 1 + \kappa' 
    \,,\quad 
    C_{--} = 1
    \,,\quad
    C_{\pm 0} = C_{0\pm} = 0
	\,,\quad
	C_{\pm\mp}=0
	\\[2mm]
	d &= \frac{1}{4\kappa'}\left[  
		C_-^2-\left( C_0^2+2C_+C_- \right) + \frac{C_+^2}{1-\kappa'}
	\right]
	- 2j(j+1)
	\,.
\end{aligned}
\end{align}
The other coefficients $C_{\pm}$ and $C_0$ can be derived from the consistency condition \eqref{eq:cnsteq} given in \cite{Ganguly:2000qy}. The derivation is tedious, but straightforward and we left further illustrations in Appendix \ref{sec:dslame}.
The complete set of solutions to \eqref{eq:cnsteq} can be categorized into three groups, as summarized in Table \ref{eq:solnce}. Qualitative discussions on this set of solutions and some examples of specific representations are provided as follows.
\begin{table}[t]
\centering
\caption{\small A set of solutions to consistency equations of QES. \label{eq:solnce}}\vspace{-3ex}
\[\begin{array}{ccccc}
	\hline
	\mbox{Group}&C_+&C_-&C_0&\alpha \Tstrut\Bstrut\\\hline\hline
	~~(\mbox{A1})~~ &\mp i(1-\kappa')& \pm i& -2j\kappa'& \mp\frac{1}{2}(1+4j)\Tstrut\Bstrut \\[2mm]
	% ~~(A2)~~&\pm i(1-\kappa')& \pm i,& -2(j+1)\kappa'& \mp\frac{1}{2}(3+4j) \Tstrut\Bstrut\vspace{1ex}\\[2mm]
	~~(\mbox{B1})~~&\mp i\left( \sqrt{1-\kappa'} - 1 + \kappa' \right)& \pm i\left( -1 + \sqrt{1-\kappa'} \right)& -(2j+1)\kappa'& \pm (2j+1)
	\Tstrut\Bstrut\\[2mm]
	~~(\mbox{B2})~~&\mp i\left( \sqrt{1-\kappa'} + 1 - \kappa' \right)& \pm i\left( 1 + \sqrt{1-\kappa'} \right)& -(2j+1)\kappa'& \mp (2j+1)
	\Tstrut\Bstrut
	\\[2mm]\hline
\end{array}\]
\end{table}

To start with, note that the group (A1) in Table \ref{eq:solnce} with $j=0$ conincides with the supersymmetric ground state. To see this is the case, for $j=0$, there are no contributions from $T^{\pm}$ and $T^0$, but from the constant term, which is $1/4$. On the other hand, we have from the right-handed side of \eqref{eq:susylamedual} that 
\begin{align}
	\frac{1}{4} = \left[ \left( \frac{1+4j}{2} \right)^2 - E_j \right]_{j=0}
\end{align}
implying the vanishing of the ground state energy. 

Generally speaking, due to the existence of the solutions in Table \ref{eq:solnce} to consistency equations, one would hence conclude that the supersymmetric is quasi-exactly solvable. In other words, the tasks of solving the differential equation for the eigenstates and eigenenergy are then translated to the problem of solving eigenvalues and eigenvectors of $(2j+1) \times (2j+1)$ matrices of the spin-$j$ representations of $sl(2)$ algebra. The explicit matrix form of the dual Hamiltonian can be found by inserting the solutions given in table \ref{eq:solnce} to the general expression \eqref{eq:tildeH}. 
However, this observation does not hold universally across all values of $\kappa'$ as some eigenvalues emerge as complex for these specific $\kappa'$ values. 
We consider some numerical investigations on the eigenvalues of the dual Hamiltonian for different sets of solutions in Table \ref{eq:solnce}.

\begin{figure}[H] 
    \centering 
    \begin{subfigure}[b]{.47\linewidth}
      \centering
      \includegraphics[width=0.95\textwidth]{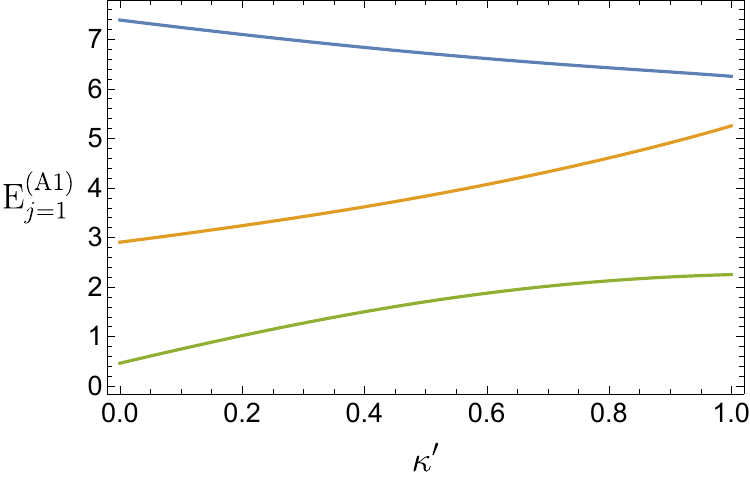}
      \caption{\small $E^{(A1)}_{j=1}$}
    \end{subfigure}
	\begin{subfigure}[b]{.47\linewidth}
      \centering
      \includegraphics[width=0.95\textwidth]{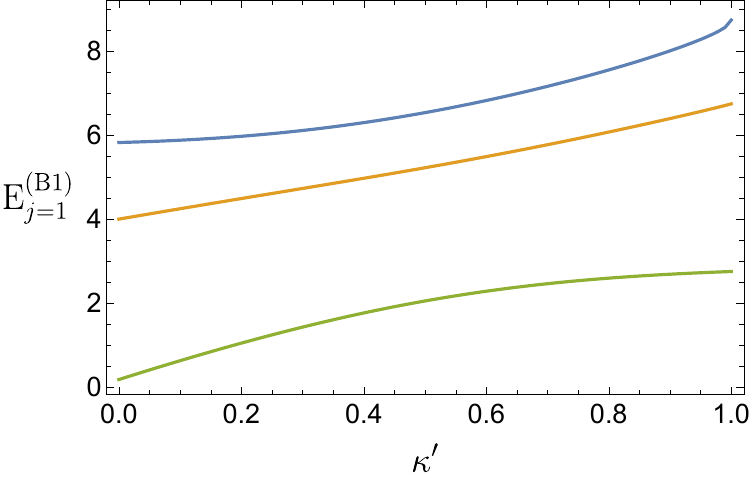}
      \caption{\small $E^{(B1)}_{j=1}$}
    \end{subfigure}
    \caption{\small Eigenenergy of group (A1) and (B1) in spin-1 representation.}
    \label{fig:LAj1}
\end{figure}

\begin{figure}[H] 
    \centering 
    \begin{subfigure}[b]{.47\linewidth}
      \centering
      \includegraphics[width=0.95\textwidth]{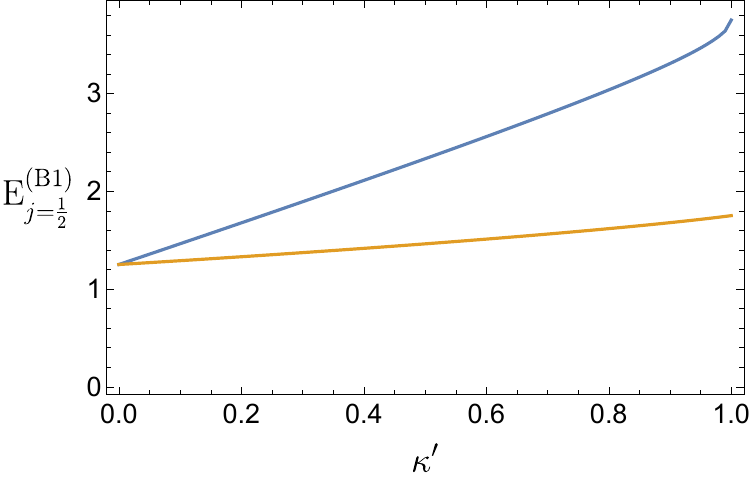}
    %   \caption{\small $E^{(A1)}_{j=1/2}$}
    \end{subfigure}
    \begin{subfigure}[b]{.47\linewidth}
        \centering
        \includegraphics[width=0.95\textwidth]{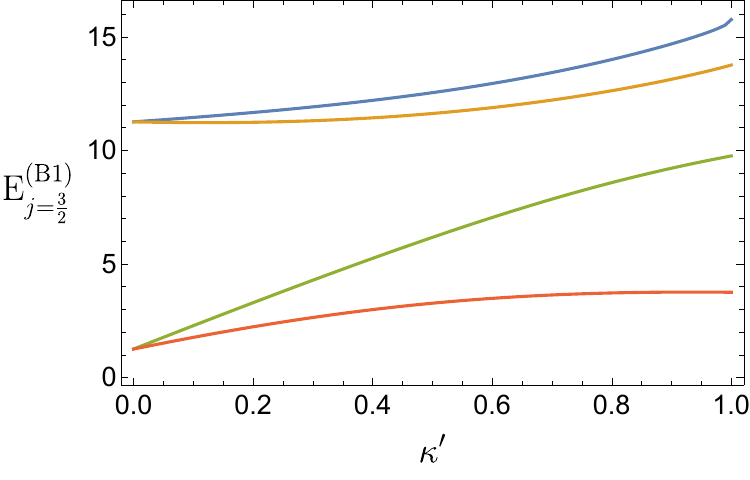}
        % \caption{\small $E^{(B2)}_{j=3/2}$}
    \end{subfigure}
	\\[2mm]
	\begin{subfigure}[b]{.47\linewidth}
      \centering
      \includegraphics[width=0.95\textwidth]{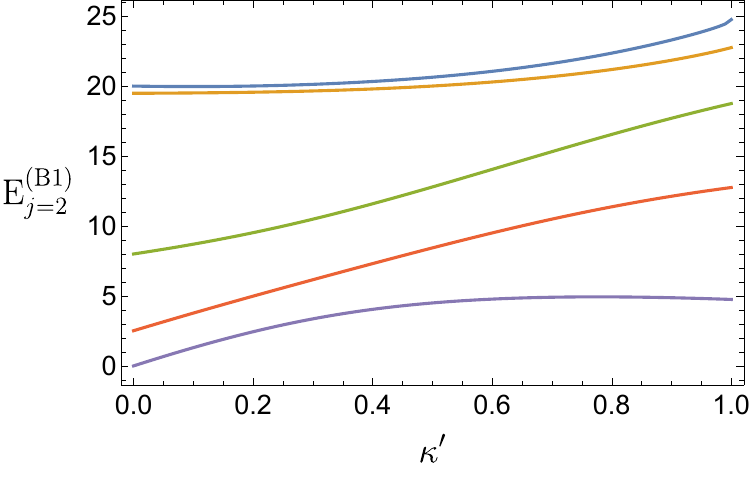}
    %   \caption{\small $E^{(B1)}_{j=2}$}
    \end{subfigure}
	\begin{subfigure}[b]{.47\linewidth}
		\centering
		\includegraphics[width=0.95\textwidth]{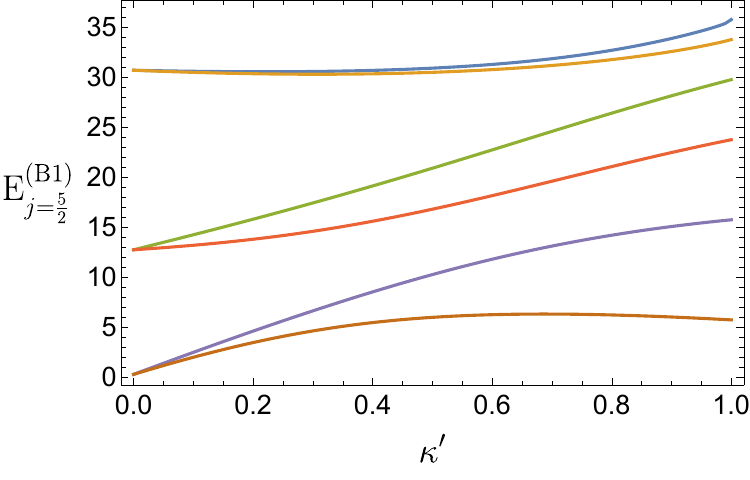}
	% \caption{\small $E^{(B1)}_{j=5/2}$}
	  \end{subfigure}
    \caption{\small Energy of eigenstates of dual Hamiltonian for $j=1/2,3/2,2,5/2$. The case of $j=1$ was presented in figure \ref{fig:LAj1}.}
    \label{fig:LAB1}
\end{figure}

Here we give the examples of quasi-exact solvability of some lower spin representations, for instance, $j=1/2$ and $1$. For the spin-$1/2$ sector, the Hamiltonian takes the forms 
\begin{subequations}
\begin{align}
	H_{j=1/2}^{(A1)} &= 
	\begin{pmatrix}\displaystyle
		\frac{3}{4}(2+\kappa') & \pm i(1-\kappa')\\[2mm] 
		\mp i & \displaystyle \frac{1}{4}(6-\kappa')
	\end{pmatrix}
	\\[2mm]
	H_{j=1/2}^{(B1)} &= 
	\begin{pmatrix}\displaystyle
		\frac{1}{4}(7+ 8\kappa'-2\sqrt{1-\kappa'}) & \pm i(1-\kappa'-\sqrt{1-\kappa'})\\[2mm] 
		\mp i\left( 1-\sqrt{1-\kappa'} \right) & \displaystyle \frac{1}{4}(7-2\sqrt{1-\kappa'})
	\end{pmatrix}
	\\[2mm]
	H_{j=1/2}^{(B2)} &= H_{j=1/2}^{(B1)}(\sqrt{1-\kappa'} \to -\sqrt{1-\kappa'})
\end{align}
\end{subequations}
with eigenvalues 
\begin{subequations}
\begin{align}
	\label{eq:e12a1}
	E_{j=1/2}^{(A1)} &= \frac{6+\kappa'}{4} \pm \frac{2-\kappa'}{2}
	% \frac{1}{4}\left( 6+\kappa' \mp 2\sqrt{\kappa^{\prime 2}+4\kappa' -4} \right)
	\\[2mm]
	\label{eq:e12b1}
	E_{j=1/2}^{(B1)} &= \frac{1}{4}\left( 
		7+4\kappa'-2\sqrt{1-\kappa'}
		\mp \sqrt{\kappa^{\prime 2}+\sqrt{1-\kappa'}(\kappa'-2)+2(1-\kappa')} 
	\right)
	\\[2mm]
	\label{eq:e12b2}
	E_{j=1/2}^{(B2)} &= E_{j=1/2}^{(B1)}(\sqrt{1-\kappa'} \to -\sqrt{1-\kappa'})
\end{align}
\end{subequations}
The energy of the group (A1) with $j=1/2$ is real for all $\kappa'$.
Regarding case (B1) and (B2), it is observed that the elements in the square roots of Eq. \eqref{eq:e12b1} and \eqref{eq:e12b2} remains non-negative for $0<\kappa'<1$.
This signifies that the system is fully quasi-exactly solvable across the entire interval for all three parametrization.
In the spin-1 sector, a slightly different observation is made as in the spin-$1/2$ scenario, where the associated energy within the group (B2) is real over a specific interval of $\kappa'$. Meanwhile, the energy for groups (A1) and (B1) remains real across the entire range of $\kappa'$. This is most effectively illustrated by figure \ref{fig:LAj1}, given that the expressions for the dual Hamiltonians are lengthy and their physical meaning is not immediately clear from their precise formulations. The energy of the eigenstates of the dual Hamiltonian for $j=1/2,1,...,5/2$ are depicted in figure \ref{fig:LAB1}.


To wrap up the current section, we comment on the supersymmetric Lam\'e model with twisted masses. By employing the same steps as outlined earlier, one can similarly derive the supersymmetric Lam\'e model. The configuration of the potential terms remains nearly identical to that of the model without twisted masses, with the primary distinction being in the associated coefficients. Namely,
\begin{align}
	\alpha^2 \to \alpha^2 + \abs{m}^2 
	\,,\quad 
	\alpha \to \alpha + m
\end{align}
in which we assume $m$ is real for simplicity. Also, we can expect that this system is Lie-algebraic and quasi-exactly solvable because the twisted mass as an additional parameter increases the degree of freedom of \eqref{eq:cnsteq}.

\section{Conclusions and outlooks}
\label{sec:conclusion}

In our study, we explored various aspects related to the Lie-algebraic deformation of the $\cp^1$ sigma model in two dimensions. We stated from generalizing  the original model endowing it with heterotic ${\cal N}=(0,2)$ and extended ${\cal N}=(2,2)$ supersymmetries. Then, we identified the hypercurrent anomaly in both scenarios. Notably, in the extended supersymmetry case, the anomaly, while being one-loop, deviates from conventional formulations found in the literature \cite{Komargodski:2010rb} due to the non-symmetric nature of the target K\"ahler manifold. We further elucidated the incorporation of twisted masses and the $\theta$ term and examined the BPS equation for instantons, focusing on its connection to the topological charge.

Moreover, we established that the second-loop contribution to the $\beta$ function in the {\em non-supersymmetric} Lie-algebraic sigma model arises from an infrared phenomenon. We used a supersymmetric regularization to substantiate our findings. We suggest that this inference  extends to higher loops, drawing a parallel with similar behaviors observed in four-dimensional ${\cal N}=1$ super-Yang-Mills theory.

In the second half of the paper, we point out the relationship between the Lie-algebraically deformed $\cp^1$ model and Lam\'e-type quantum mechanics, achieved through the Scherk-Schwarz dimensional reduction technique. This result is then further generalized to the case of {\ntwoo} supersymmetry. Upon certain additional requirements, the supersymmetric quantal problem obtained in this way proves to be quasi-exactly solvable. Further research into the link between two-dimensional Lie-algebraic models and quasi-exactly solvable quantum mechanics could focus on the deformation of 2D sigma models via specific potentials and the generalization of Lam\'e quantum mechanics, including the study of the associated Lam\'e equation.

\section*{Acknowledgments}
M.S. is grateful to A. Turbiner and O. Gamayun for useful discussions.  
This work is supported in part by DOE grant DE-SC0011842 and the Simons Foundation Targeted Grant 920184 to the Fine Theoretical Physics Institute.

\section*{Appendices}
\appendix

\section{Conventions on elliptic integrals and Jacobi elliptic function}
\label{sec:jef}

In this appendix, we provide a summary of the definitions and several useful identities of Jacobi elliptic functions that are used in the main text. Further properties can be found in, for example, \cite{NIST:DLMF,abramowitz+stegun}. First, the elliptic integral of the first kind is defined as 
\begin{align}
	F(\kappa|\phi) \equiv \int_{0}^{\phi} \frac{\dd{t}}{\sqrt{1-\kappa\sin^2{t}}}
	\,,
\end{align}
where the complete elliptic integral $K(\kappa) = F(\kappa|\pi/2)$.
To define Jacobi elliptic functions, it suffices to consider the inverse of the incomplete integral $F(\kappa|\phi)$ 
\begin{align}
	\phi \equiv \operatorname*{a.m.}(F|\kappa) \,.
\end{align}
Then the Jacobi elliptic functions are represented as 
\begin{align}
\begin{aligned}
	&\sn(F|\kappa) \equiv \sin{\phi} 
	\,,\quad 
	\cn(F|\kappa) \equiv \cos{\phi}
	\,,\quad 
	\dn(F|\kappa) \equiv \sqrt{1-\kappa\sin^2{\phi}}
	\\[2mm]
	&\sd(F|\kappa) \equiv \frac{\sin{\phi}}{\sqrt{1-\kappa\sin^2{\phi}}} 
	\,,\quad 
	\cd(F|\kappa) \equiv \frac{\cos{\phi}}{\sqrt{1-\kappa\sin^2{\phi}}} 
	\,.
\end{aligned}
\end{align}
with the elliptic modulus $\kappa \in [0,1)$ and $\kappa'\equiv 1-\kappa$.

The following identities are used in discussion of the Lam\'e systems, in particular, in Sec. \ref{sec:lame} for the dual transformations,
\begin{align}
\begin{aligned}
    \kappa \sn^2(\theta,\kappa) &= 1 - \kappa'\sn^2(\theta',\kappa')\,,
    \\[2mm]
    \sqrt{\kappa} \cn(\theta,\kappa)\dn(\theta,\kappa) &= i\kappa'\sn(\theta',\kappa')\cn(\theta',\kappa')\,.
\end{aligned}
\end{align}

\section{Derivation of anomalous axial $U(1)$ currrent}
\label{sec:acrt}

In this section, we outline the derivation of the connection between the divergence of the anomalous axial current and the theta term of the deformed $\cp^1$ model. 

Parallel to the discussion in \cite{FS}, one has that 
\begin{align}\label{eq:u1current}
    \d_{\mu}\left( G\bpsi{}\gamma^{\mu}\gamma_5\psi \right) 
    &= 
    \d_{\mu}\left( \bar{\chi}\gamma^{\mu}\gamma_5\chi   \right)
    = 2i\Tr\gamma_5 f\left( \frac{\cancel{D}^2}{\Lambda^2} \right)
\end{align} 
where $f(x)$ is the regularization function with cutoff $\Lambda$ such that 
\begin{align*}
	f(0) = 1 
	\qand
	\lim_{x\to\infty}f(x)=0 \,.
\end{align*}
Here we consider the vielbein to decompose the metric, say,
\begin{align}
    \bar{\chi} = \sqrt{G}\bpsi{}
    \,,\quad
    \chi = \sqrt{G}\psi
	\,,
\end{align}
and the covariant derivative $D_{\mu}$ under this frame is 
\begin{align*}
	D_{\mu} \equiv \frac{1}{2} \left( \Gamma\d_{\mu}\varphi - \overline{\Gamma}\d_{\mu}\bphi{} \right)
	\,.
\end{align*}
Consequently, the trace turns out to be 
\begin{align}\label{eq:axial}
	\Tr\gamma_5 f\left( \frac{\cancel{D}^2}{\Lambda^2} \right) 
    &= \Lambda^2\tr\int\frac{\dd^{2}{k}}{(2\pi)^2}\gamma_5f\left(  
        -k^2 + \frac{2i(k \cdot D)}{\Lambda} + \frac{D^2}{\Lambda^2}
        -\frac{1}{4\Lambda^2}[\gamma^{\mu},\gamma^{\nu}]R_{(2),\mu\nu}
    \right)
	\nonumber\\[2mm]
	&\to \frac{1}{2\pi} R_{1\bar{1}} \epsilon^{\mu\nu}\d_{\mu}\bphi{}\d_{\nu}\varphi \,.
\end{align}
To get the second line of \eqref{eq:axial}, we take the limit $\Lambda \to \infty$ and employ the commutation relation 
\begin{align*}
	[D_{\mu},D_{\nu}] &= 
	% -R_{1\bar{1}}\left( 
    %     \d_{\mu}\bphi{}\d_{\nu}\varphi
    %     - \d_{\nu}\bphi{}\d_{\mu}\varphi
    % \right)
    % \equiv 
	-R_{(2),\mu\nu}
\end{align*}
in which 
\begin{align}
	R_{(2),\mu\nu} \equiv 
	R_{1\bar{1}}\left( 
        \d_{\mu}\bphi{}\d_{\nu}\varphi
        - \d_{\nu}\bphi{}\d_{\mu}\varphi
    \right)
	\,.
\end{align}
Therefore, combining \eqref{eq:u1current} and \eqref{eq:axial}, we have 
\begin{align}
    \d_{\mu}\left( G\bpsi{}\gamma^{\mu}\gamma_5\psi \right)
    = -\frac{i}{\pi} R_{1\bar{1}}\epsilon^{\mu\nu}\d_{\mu}\varphi\d_{\nu}\bphi{}
	\,.
\end{align}
Together with \eqref{scal}, we arrive at \eqref{eq:j5dq}.

\section{Comparison between deformed \boldmath{$\cp^1$} quantum mechanics and Lam\'e quantum mechanics}
\label{sec:dcpvslame}

Here the distinction between different dimensional reduction scenarios are detailed. In particular, we compare the resultant quantum mechanics from the Kaluza--Klein (KK) and the Scherk-Schwarz reductions. 

In the Kaluza--Klein framework, one assumes the spacetime dependence of the field to be 
\begin{align}
	\varphi(t,z) = \sum_{n=0}^{\infty}\varphi_{(n)}(t)\exp(i\frac{2\pi nz}{L})
	\,,\quad
	\bphi(t,z) = \sum_{n=0}^{\infty}\bphi_{(n)}(t)\exp(-i\frac{2\pi nz}{L})\,.
\end{align}
As substituting this into \eqref{3three}, integrating along the $z$-direction, and keeping only the lowest mode, we have 
\begin{align}\label{eq:Ldefmcp1}
	\cL_{KK} = G\dot{\varphi}\dot{\bphi}
	= \dot{\theta}^2 + \sn^2(\theta|\kappa)\dot{\alpha}^2
\end{align}
from which we can see that there will be some additional pieces in the deformed $\cp^1$ Hamiltonian, comparing with the Lam\'e equation. Indeed, according to Eq. \eqref{eq:Ldefmcp1}, the Hamiltonian of the deformed $\cp^1$ model is 
\begin{align}
	H_{d\cp^1} = -\frac{1}{4}\Delta 
	= -\frac{1}{4}\left[ 
		\dv[2]{\theta} 
		+ \frac{1}{\sn(\theta|\kappa)}\dv{\theta}
		+ \frac{1}{\sn^2(\theta|\kappa)}\dv[2]{\alpha}
	\right]
\end{align}
where $\Delta$ is the Laplace operator. Note that \eqref{eq:Ldefmcp1} is a \emph{two-dimensional} quantum mechanical system while the Lam\'e model \eqref{eq:lameH} is of one dimension which depends only on $\theta$. Thus even we consider the zero-angular momentum sector of the Hilbert space (i.e. $\pdv*{\Phi}{\alpha}=0$), there is still an additional linear contribution in $\theta$ in the deformed $\cp^1$ Hamiltonian than the Lam\'e one. If one keeps a higher mode rather than the lowest one, it will introduce an extra mass term, but can still do nothing with eliminating the linear differential part in $\theta$. The further discussion on the deformed $\cp^1$ quantum mechanics derived from the KK reduction can be found in \cite{CS}. In fact, the similar issue between the $\cp^1$ model and the sine-Gordon model was discussed in \cite{Fujimori:2016ljw} from the perspective of resurgence analysis. 

\section{Details of Lie-algebraic features of supersymmetric Lam\'e quantum mechanics}
\label{sec:dslame}
The sufficient and necessary condition for rendering the system quasi-exactly solvable is that there exists non-trivial solutions $(C_{\pm},C_0,\alpha)$ to the equations of consistency\footnote{These consistent equations arise from the change of variables from $\xi$ to $\theta'$.} given in \cite{Ganguly:2000qy}, namely,
\begin{subequations}\label{eq:cnsteq}
\begin{align}\label{eq:ce1}
    &\kappa'j(j+1) - \frac{C_0}{2}(2j+1) + \frac{1}{4\kappa'}\left[ C_0^2-(C_+-C_-)^2 \right] = \alpha^2\kappa'\,,
    \\[2mm]
	\label{eq:ce2}
    &\frac{1}{2\kappa'}\left( C_+ -C_- \right)\left[ \kappa'(2j+1) - C_0 \right] = i\alpha\kappa'\,,
    \\[2mm]
	\label{eq:ce3}
    &\frac{1}{2\kappa'}\left[ C_+ - (1-\kappa')C_- \right]\left[ \kappa'(2j+1) + C_0 \right] = 0\,,
    \\[2mm]
	\label{eq:ce4}
    &\kappa'j(j+1) + \frac{C_0}{2}(2j+1) + \frac{1}{4\kappa'}\left[ C_0^2-\frac{(C_+ - (1-\kappa')C_-)^2}{1-\kappa'} \right]
    = 0\,.
\end{align}
\end{subequations}
For group (A1), it corresponds to taking $C_+ = (1-\kappa')C_-$ while taking $C_0 = \kappa'(2j+1)$ is related to group (B1) and (B2).

As reducing to the (bosonic) Lam\'e case, we have the vanishing right-handed side of \eqref{eq:ce2}. The solution to this reduced case is $C_\pm = 0,C_0 = -2jk,\alpha = 2j(1+2j)$ as claimed in \cite{dunne} since $j$ is a semi-integer.

{\small\providecommand{\href}[2]{#2}
\providecommand{\doihref}[2]{\href{#1}{#2}}
\providecommand{\arxivfont}{\rm}

}

\end{document}